\documentclass[aps,prc,superscriptaddress,showpacs,floatfix]{revtex4}

\usepackage{amssymb}
\usepackage{amsmath}
\usepackage{graphicx}
\usepackage[normalem]{ulem}
\usepackage{multirow}
\usepackage[usenames]{color}
\usepackage{bm}
\usepackage{verbatim}
\begin{document}

\title{
Two-particle angular correlations in heavy ion collisions from a multiphase transport model
}

\author{Liu-Yao Zhang}
\affiliation{Shanghai Institute of Applied Physics, Chinese Academy of Sciences, Shanghai 201800, China}
\affiliation{University of Chinese Academy of Sciences, Beijing 100049, China}
\author{Jin-Hui Chen}
\affiliation{Institute of Modern Physics and Key Laboratory of Nuclear Physics and Ion-beam Application (MOE), Fudan University, Shanghai 200433, China}
\author {Zi-Wei Lin}
\affiliation{Department of Physics, East Carolina University, Greenville, North Carolina 27858, USA}
\affiliation{Key Laboratory of Quarks and Lepton Physics (MOE) and Institute of Particle Physics, Central China Normal University, Wuhan 430079, China}
\author{Yu-Gang Ma}
\affiliation{Shanghai Institute of Applied Physics, Chinese Academy of Sciences, Shanghai 201800, China}
\affiliation{Institute of Modern Physics and Key Laboratory of Nuclear Physics and Ion-beam Application (MOE), Fudan University, Shanghai 200433, China}
\author{Song Zhang}
\affiliation{Institute of Modern Physics and Key Laboratory of Nuclear Physics and Ion-beam Application (MOE), Fudan University, Shanghai 200433, China}
\date{\today}

\begin{abstract}
We extend our earlier study on two-particle angular correlations in $pp$ collisions at low transverse momentum ($p_T$) to $p$-Pb, Pb-Pb and Au-Au collisions at RHIC and LHC energies. We mainly use the string melting version of a multiphase transport model with improved quark coalescence for this study. We start from the analysis of $\pi^{\pm}$, $K^{\pm}$ and $p$($\bar{p}$) $p_T$ and rapidity distributions at different centralities. We then focus on two-particle angular correlations in $p$-Pb collisions at $\mathrm{\sqrt{s_{NN}} = 5.02}$ TeV and Pb-Pb collisions at $\mathrm{\sqrt{s_{NN}}= 2.76}$ TeV. For $p$-Pb collisions, a near side depression in the angular correlation is observed for low $p_T$ proton pairs and $\Lambda$ pairs but not for pion pairs or kaon pairs, similar to our earlier finding for $pp$ collisions at $\mathrm{\sqrt{s}= 7}$ TeV. This is also the case for very low multiplicity Pb-Pb and Au-Au collisions. We also find that parton interactions and the improved quark coalescence are mainly responsible for the depression feature in baryon pair angular correlations. 
However, no such baryon-baryon anti-correlations are observed in Pb-Pb and Au-Au collisions at higher multiplicities. Therefore our results suggest that low $p_T$ baryon-baryon angular anti-correlations have a strong multiplicity dependence.  
\end{abstract}
\maketitle

\section{Introduction}
The scientific goal of ultra-relativistic heavy-ion collisions at Relativistic Heavy Ion Collider (RHIC) and at Large Hadron Collider (LHC) is to study the properties of a deconfined and chiral-restored state of matter: the quark-gluon plasma (QGP)~\cite{Shuryak.E.V:1978,I.Arsene:2005,B.B.Back:2005,J.Adams:2005,K.Adcox:2005,Chen:2018}, under extreme temperature and energy density. Analysis of multi-particle angular correlations is a powerful tool in exploring the properties of QGP and the underlying mechanism of particle production in hot QCD matter~\cite{M.Aaboud:2017,C.Aidala:2017,L.Adamczyk:2015,C.Adare:2015,L.Adamczyk:2014,S.Chatrchyan:2011,V.Khachatryan:2010,B.Alver:2010}. The correlations are often presented in the $\Delta\eta$-$\Delta\phi$ space, where $\Delta\eta$ is the pseudorapidity difference and $\Delta\phi$ is the azimuthal angle difference between the two particles in the pair. For example, in $pp$ collisions at LHC energies~\cite{V.Khachatryan:2010} a narrow peak around ($\Delta\eta$, $\Delta\phi$) $\sim$ (0,0) is observed for pairs having 1$<p_T<$3 GeV/c, which is thought to be coming from the showering and hadronization of the leading parton. On the away-side ($\Delta\phi \sim\pi$), a long-range structure in $\Delta\eta$ represents correlations from the recoiling parton. In heavy-ion collisions for pairs having the similar $p_T$ range, in addition to the jet-like correlations, a pronounced near-side ($\Delta\phi \sim$0) collimation extending over a long range in $\Delta\eta$ is observed, the so-called "ridge" phenomenon~\cite{S.Chatrchyan:2011}. The phenomenon is even observed in $pp$ collisions at high multiplicities~\cite{V.Khachatryan:2010}, which has attracted lots of discussions on the physics origin of this surprising behavior in small systems, such as collective flow effects (see Ref.~\cite{Dusling:2016} for a recent review). So far no conclusive explanation has been reached~\cite{Aidala:2019,Mace:2018}, and more theoretical studies and experimental data in small systems are needed.   

The experimental data in $pp$ collisions at $\mathrm{\sqrt{s}= 7}$ TeV shows a clear near-side depression of baryon-baryon angular correlations at low $p_T$, which provides new inputs to the modification of particle production mechanism or the fragmentation functions in Monte Carlo models~\cite{J.Adam:2017}. Recently we find that a multi-phase transport  (AMPT) model with new quark coalescence~\cite{Y.C.He:2017} is able to qualitatively describe the angular anti-correlations of low $p_T$ baryon-baryon pairs observed in data~\cite{L.Y.Zhang:2018}. Compared to other Monte Carlo models~\cite{J.Adam:2017}, our results~\cite{L.Y.Zhang:2018} suggest that quark coalescence and parton scatterings in the AMPT model are essential to describe the correlation features in $pp$ collisions at LHC energies. Now we extend the study to larger systems such as $p$-Pb and Pb-Pb collisions.

In this paper, we explore in detail the low $p_T$ baryon-baryon angular correlations in $p$-Pb and Pb-Pb collisions at LHC energies as well as Au-Au collisions at RHIC energies. We first study the $p_T$ spectra and $dN/dy$ distributions of identified particles in $p$-Pb and Pb-Pb collisions from the string melting AMPT model to check the improvement of the new quark coalescence model in comparison with the original AMPT model. We then study $\pi$-$\pi$, K-K, $p$-$p$, and $\Lambda$-$\Lambda$ angular correlations in $p$-Pb collisions at $\mathrm{\sqrt{s_{NN}}= 5.02}$ TeV and Pb-Pb collisions at $\mathrm{\sqrt{s_{NN}}= 2.76}$ TeV  with different values for the parton cross section and hadron cascade evolution time. We also study Au-Au collisions at $\mathrm{\sqrt{s_{NN}}}$ = 7.7 and 200 GeV to address the dependences on the collision energy and system size.

\section{A BRIEF DESCRIPION OF the AMPT MODEL}
The AMPT model~\cite{Z.W.Lin:2005} is a transport model consisting of four main components: the fluctuating initial condition, partonic rescatterings, hadronization, and hadronic interactions. The initial condition includes the spatial and momentum distribution of minijet partons and soft string excitations, which are based on the heavy-ion jet interaction generator (HIJING) model \cite{X.N.Wang:1991}, an extension of the PYTHIA model \cite{Phthia}. In the string melting version \cite{Z.W.Lin:2002} of the AMPT model, both excited strings and minijet partons are decomposed into partons. Scatterings among the partons, which are produced from initial nucleon-nucleon interactions, are modeled by Zhang's parton cascade (ZPC) \cite{B.Zhang:1998}, which includes two-body scatterings with the cross section based on the perturbative QCD calculation using a screening mass. After parton interactions stop, quark coalescence~\cite{Z.W.Lin:2002} is used to model the hadronization by combining nearby partons into hadrons. The dynamical evolution of the hadronic phase is subsequently described by an extended version of a relativistic transport (ART) model \cite{B.A.Li:1995} including baryon-baryon, baryon-meson, and meson-meson elastic and inelastic scatterings~\cite{Z.W.Lin:2005}. 

It has been found~\cite{Z.W.Lin:2014} that the string melting AMPT model, with suitable choice of a few key input parameters, can reasonably describe 
the particle yields, transverse momentum spectra, and momentum anisotropies of low $p_T$ particles in high energy heavy ion collisions. 
The key parameters include the Lund string fragmentation parameters $a = 0.30, b = 0.15$/GeV$^2$, the strong coupling constant $\alpha_s = 0.33$, and the parton cross section of $\sim 3$ mb~\cite{Z.W.Lin:2014,G.L.Ma:2016}. We use these input parameters and a parton cross section of 3 mb in this study. Another significant improvement is on the quark coalescence component~\cite{Y.C.He:2017,Y.C.He:2018}. The original AMPT model forces the numbers of mesons, baryons, and antibaryons in an event to be separately conserved through the quark coalescence process; while only the net-baryon number needs to be conserved, which is the case in the new quark coalescence model~\cite{Y.C.He:2017}. Indeed, clear improvements on describing heavy ion collisions have been achieved using the AMPT model with this new quark coalescence~\cite{Y.C.He:2017,Y.C.He:2018,L.Y.Zhang:2018,X.H.Jin:2018,Liu:2017,X.H.Jin:2019,C.Zhang:2019}. For this study, we use the string melting version of AMPT model (v2.31t1) that has incorporated the new quark coalescence~\cite{ampt}. We denote this model as AMPT newCoal in the following and denote the original AMPT results as AMPT oldCoal.

\section{Results \& discussions}

\begin{figure}[!htb]
	\centering
	\includegraphics[scale=0.25]{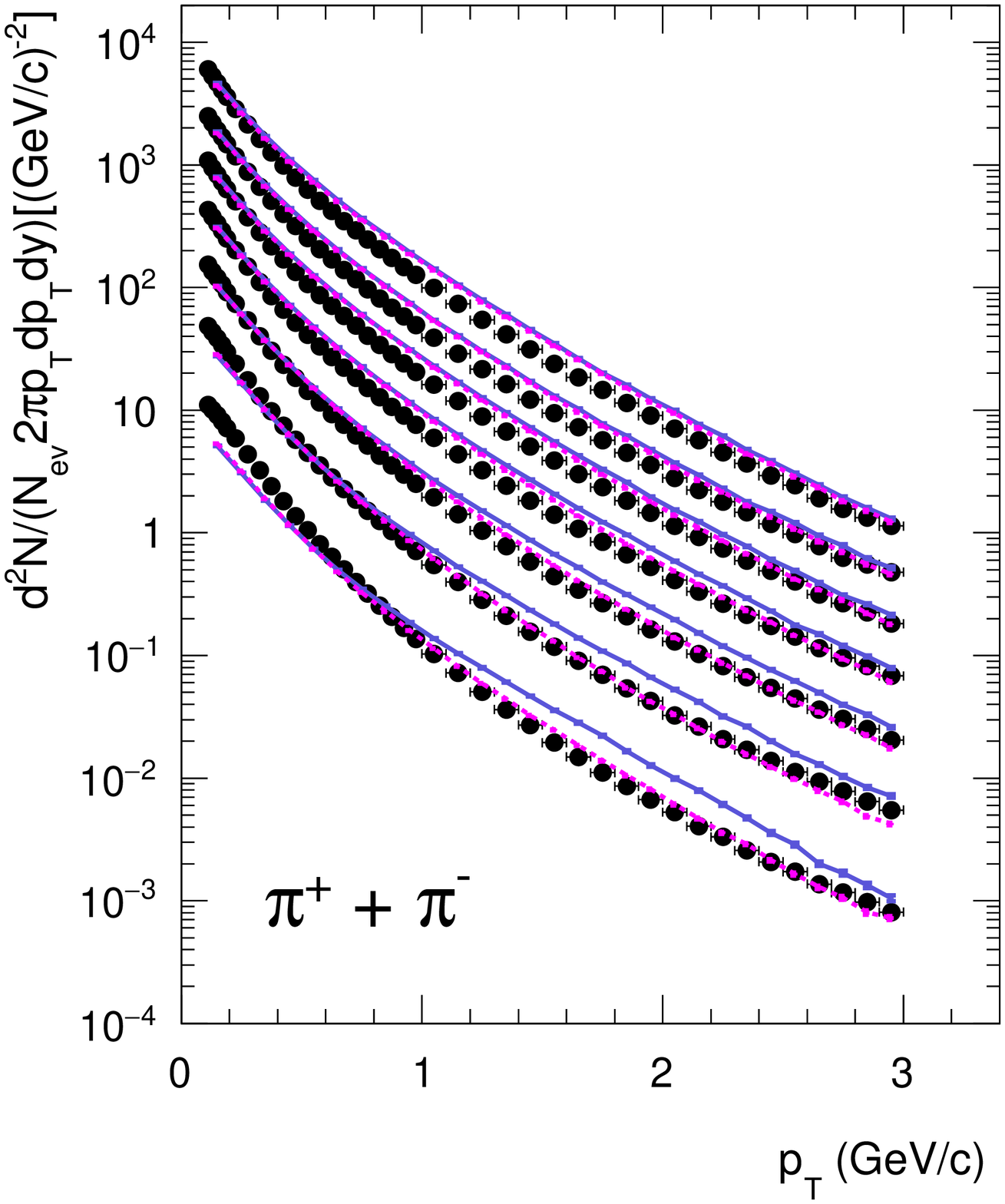}
	\includegraphics[scale=0.25]{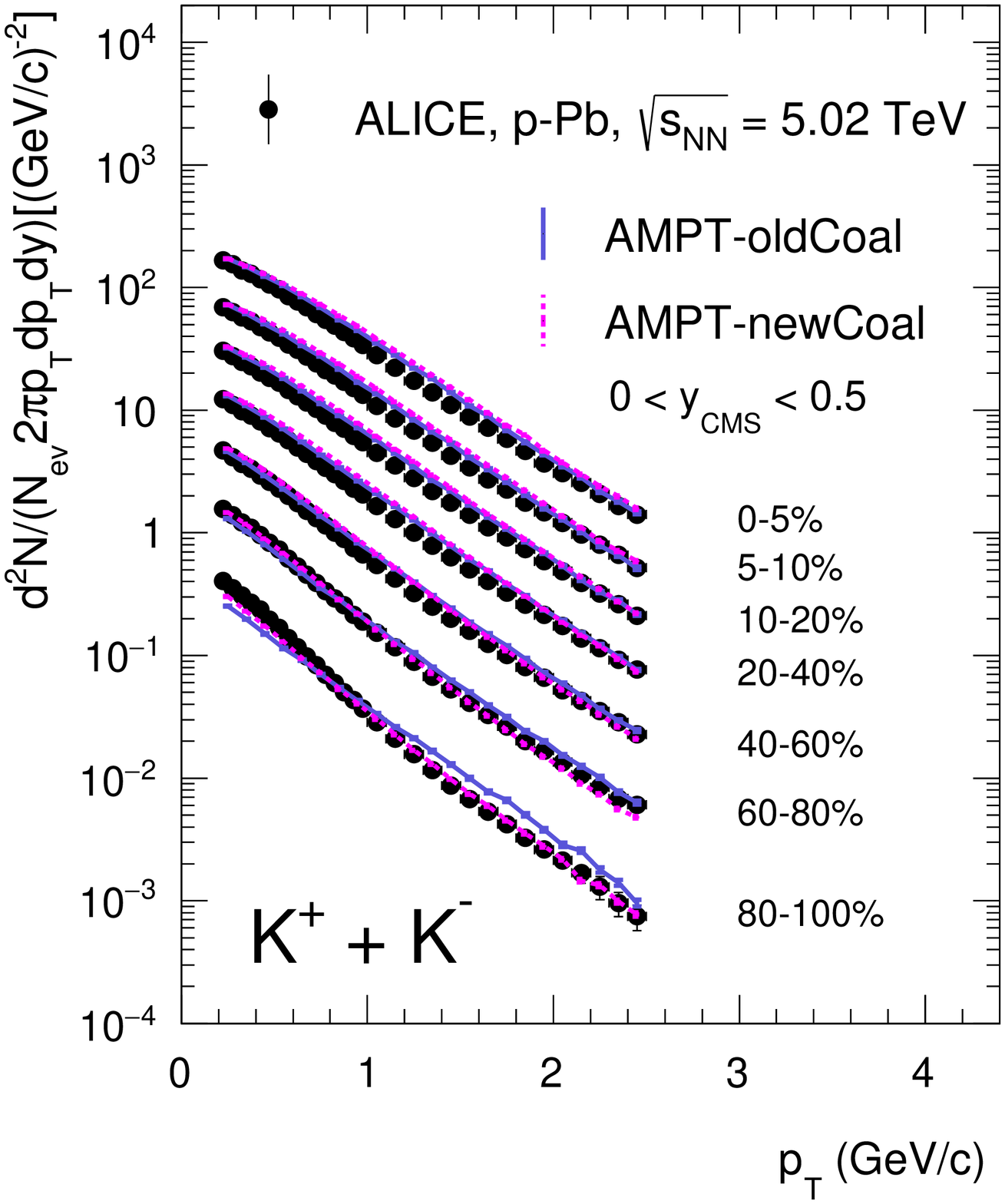}
	\includegraphics[scale=0.25]{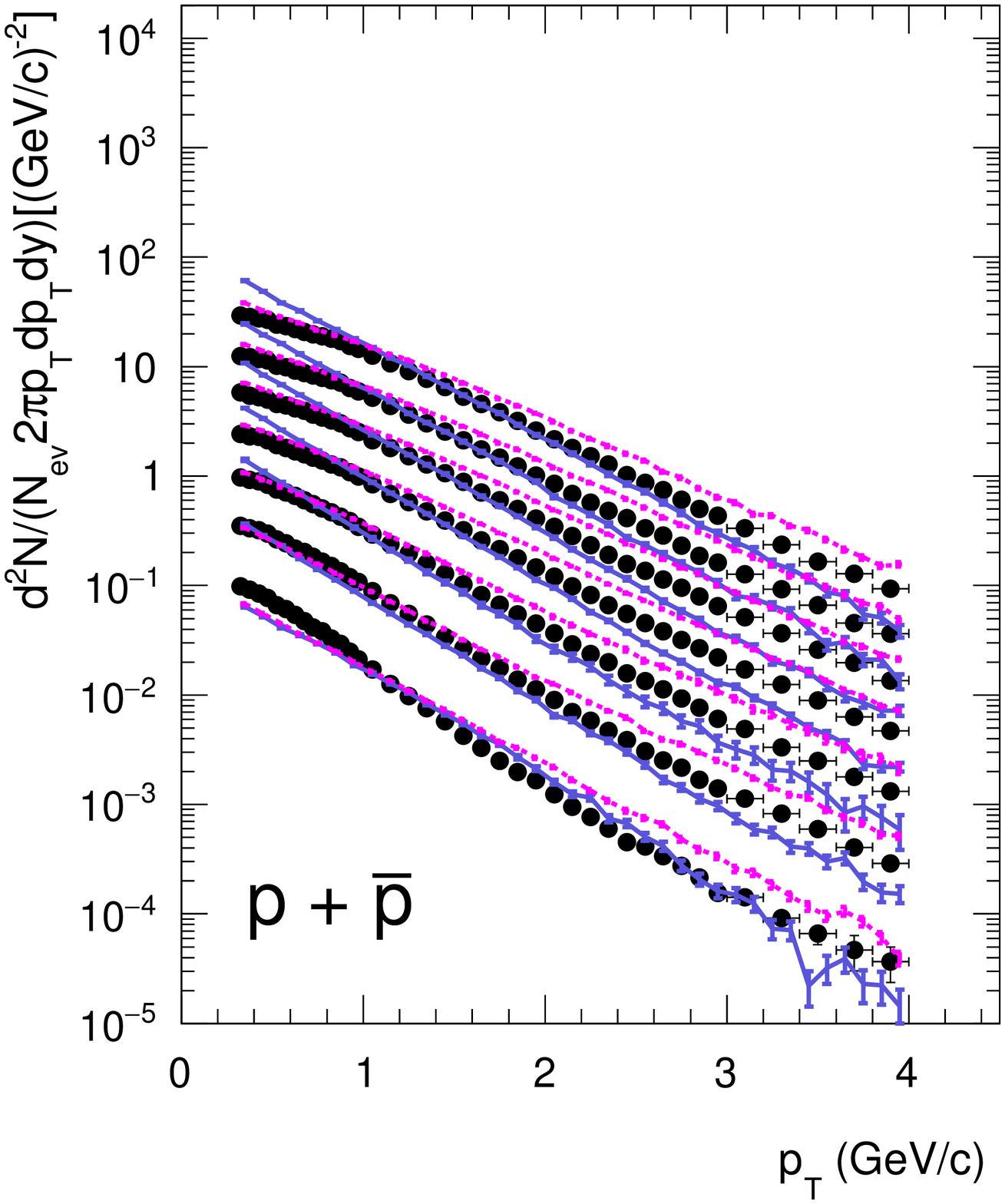}\\
	\caption{(Color online) The $p_{T}$ distributions of $\pi^{\pm}$ (left panel), $K^{\pm}$ (middle panel) and $p$($\bar{p}$) (right panel) in $p$-Pb collisions at $\mathrm{\sqrt{s_{NN}} = 5.02}$ TeV as a function of $p_{T}$ at different centralities. Solid curves represent results from the AMPT model with old quark coalescence, while dotted curves are results with the new quark coalescence. Solid circles are experimental data~\cite{B.Abelev:2014}. }
	\label{charged_particle_pt_spectra_pPb}
\end{figure}

\begin{figure}[!htb]
	\centering
	\includegraphics[scale=0.27]{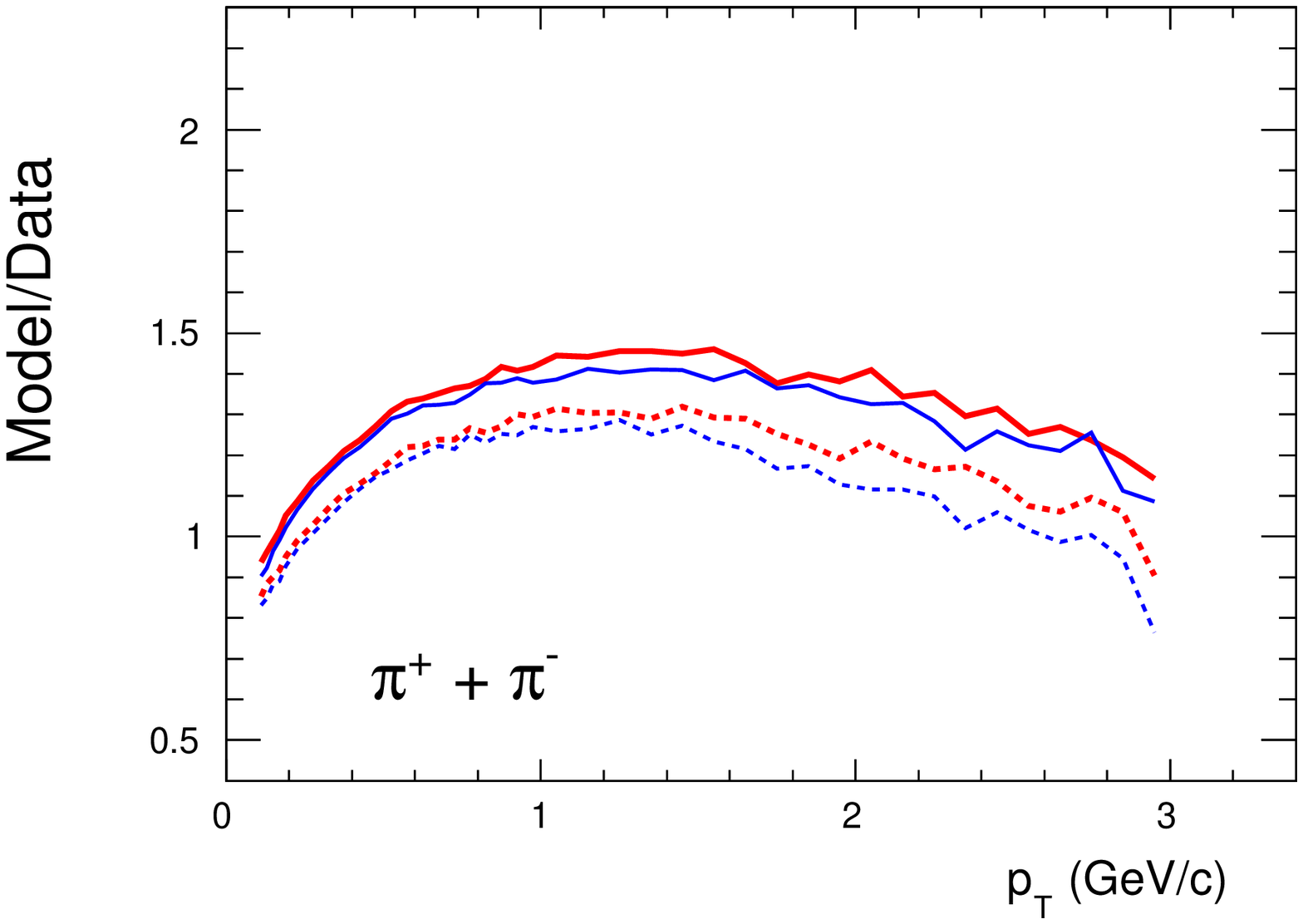}
	\includegraphics[scale=0.27]{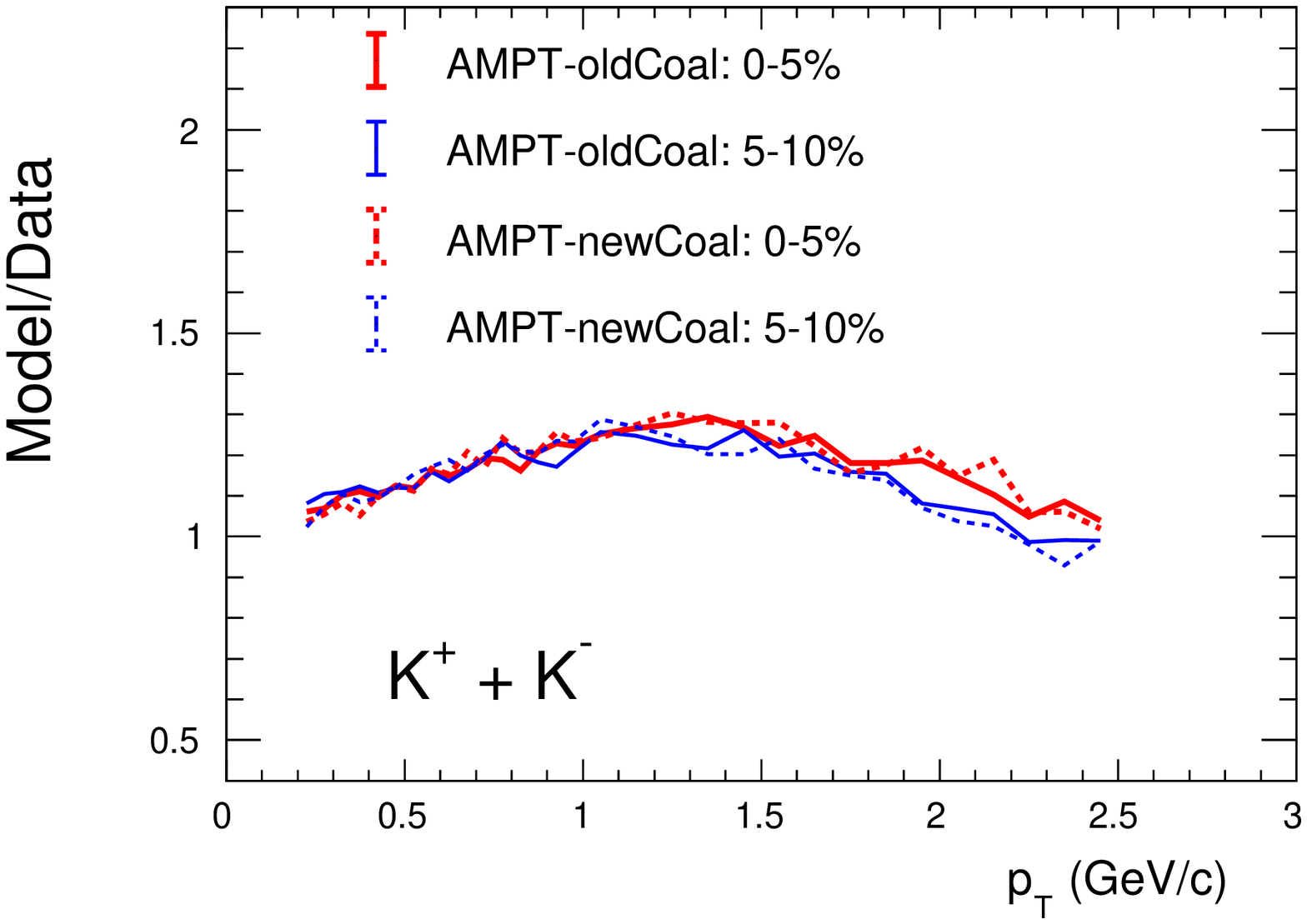}
	\includegraphics[scale=0.27]{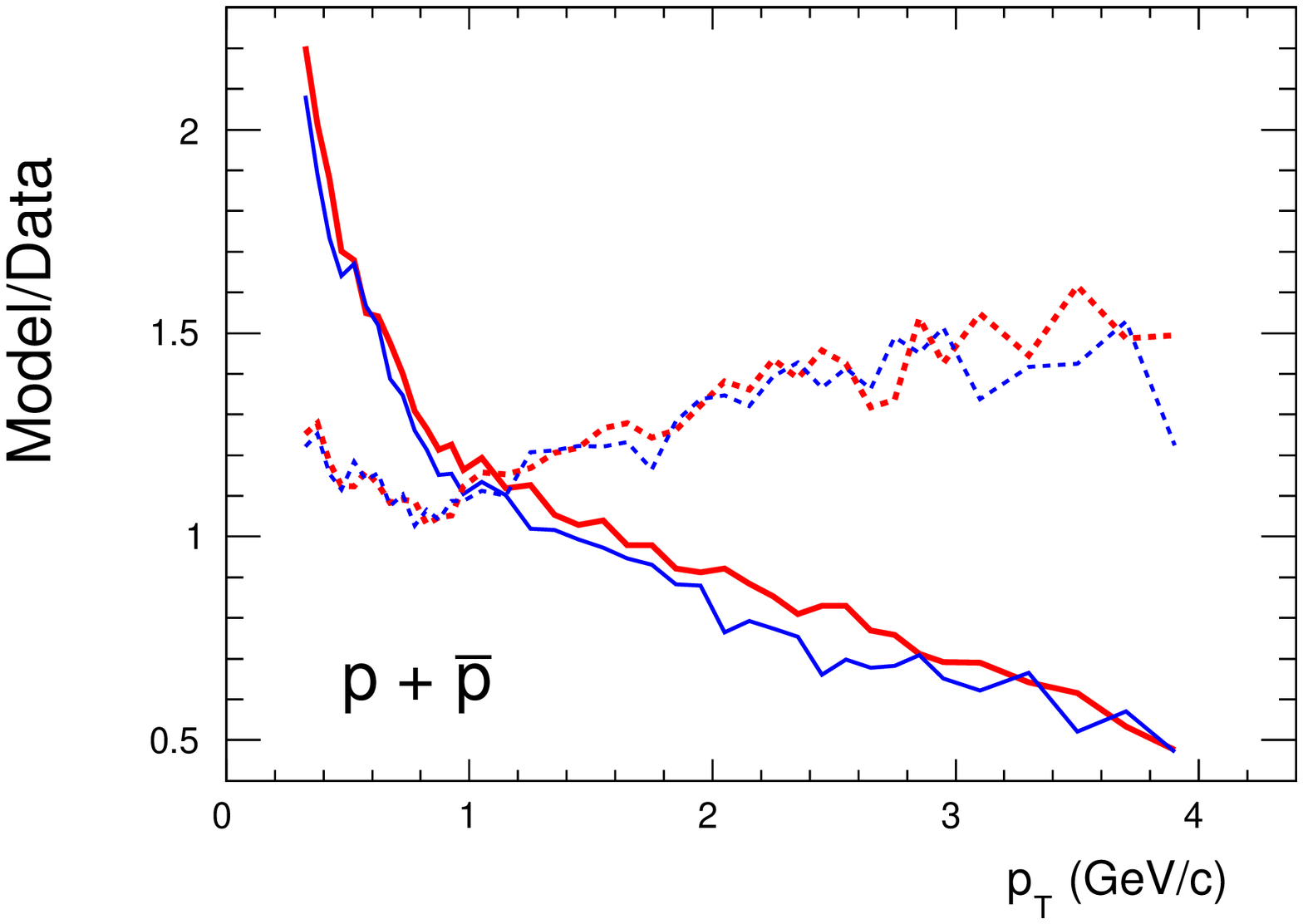}\\
	\caption{(Color online) The ratio of AMPT model results to the experimental data in central $p$-Pb collisions. Solid curves are results from AMPT with old quark coalescence, and dotted curves represent the new quark coalescence.}
	\label{Ratio_model_vs_data_pPb}
\end{figure}

\begin{figure}[!htb]
	\centering
	\includegraphics[scale=0.25]{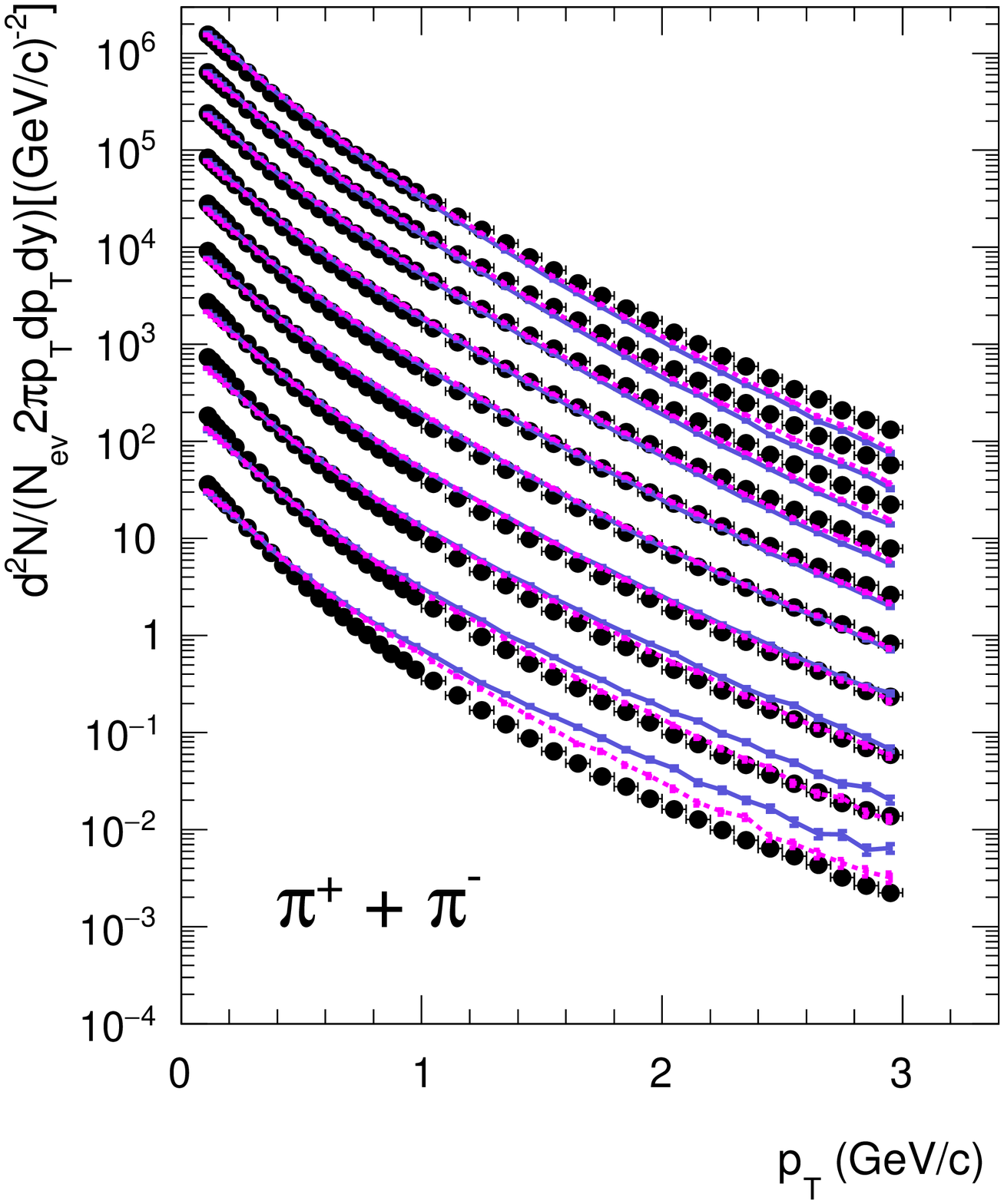}
	\includegraphics[scale=0.25]{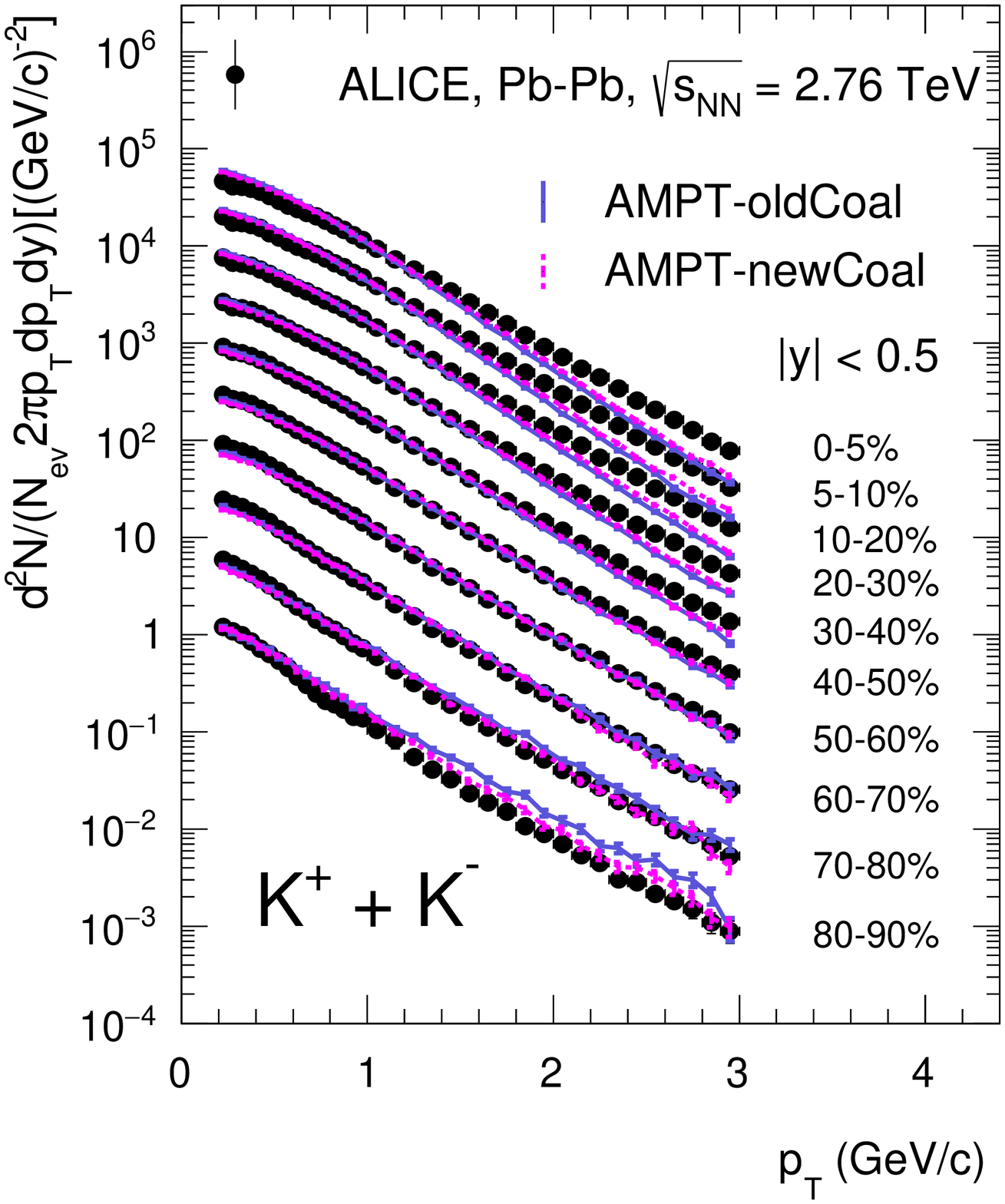}
	\includegraphics[scale=0.25]{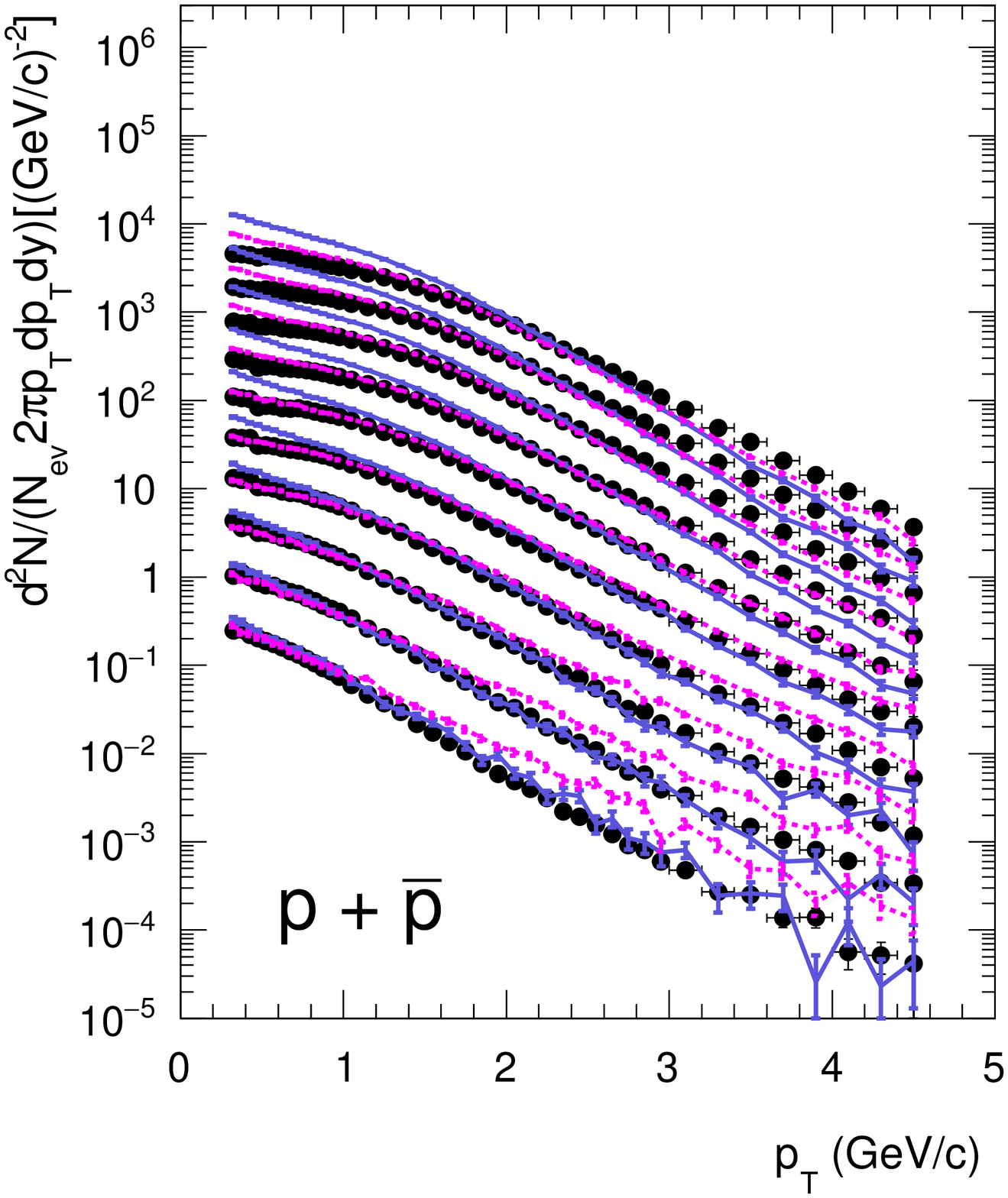}
	\caption{(Color online) The $p_{T}$ distributions of $\pi^{\pm}$ (left panel), $K^{\pm}$ (middle panel) and $p$($\bar{p}$) (right panel) in Pb-Pb collisions at $\mathrm{\sqrt{s_{NN}} = 2.76}$ TeV. Solid points are experimental data from Ref.~\cite{B.Abelev:2013}.}
	\label{charged_particle_pt_spectra_PbPb}
\end{figure}

\begin{figure}[!htb]
	\centering
	\includegraphics[scale=0.27]{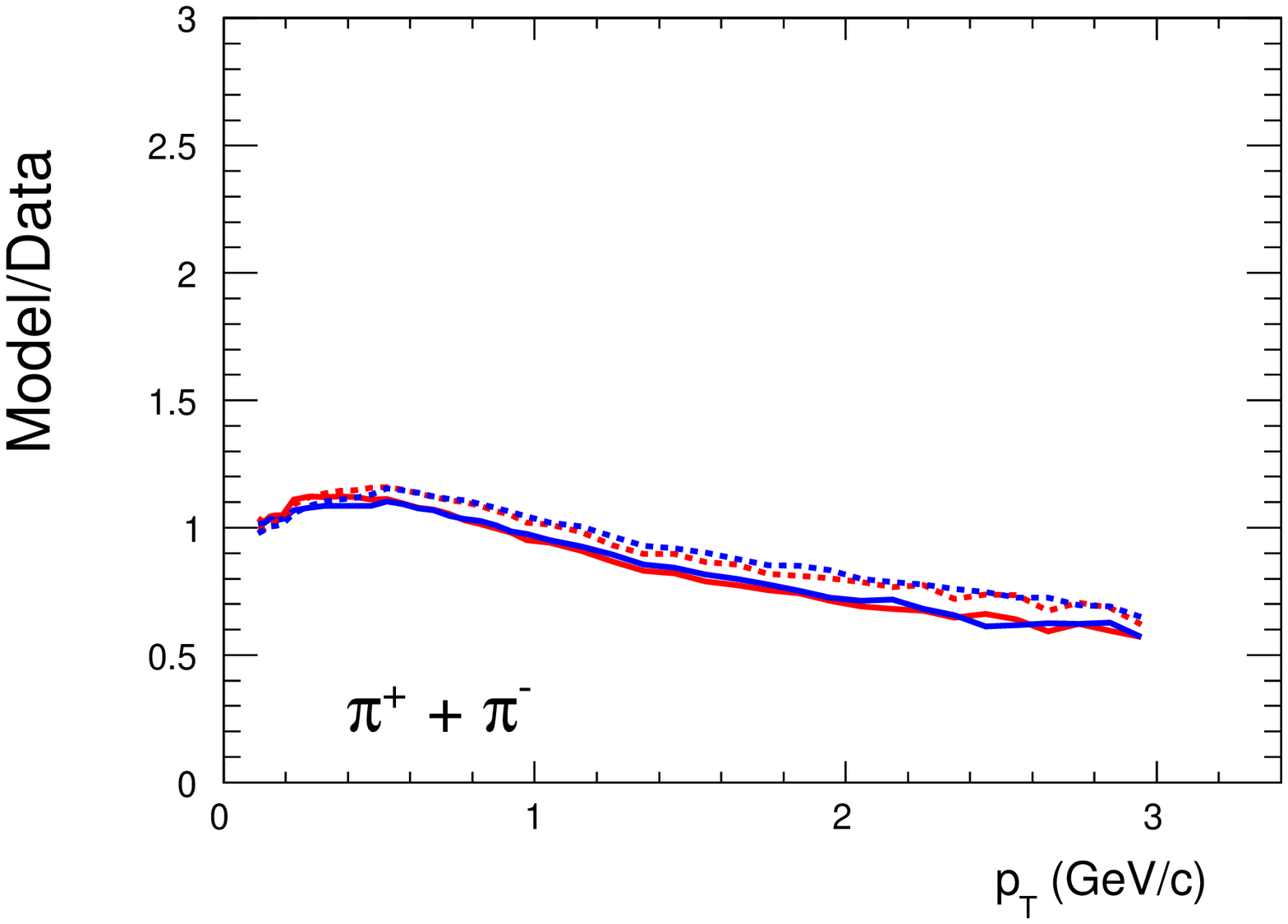}
	\includegraphics[scale=0.27]{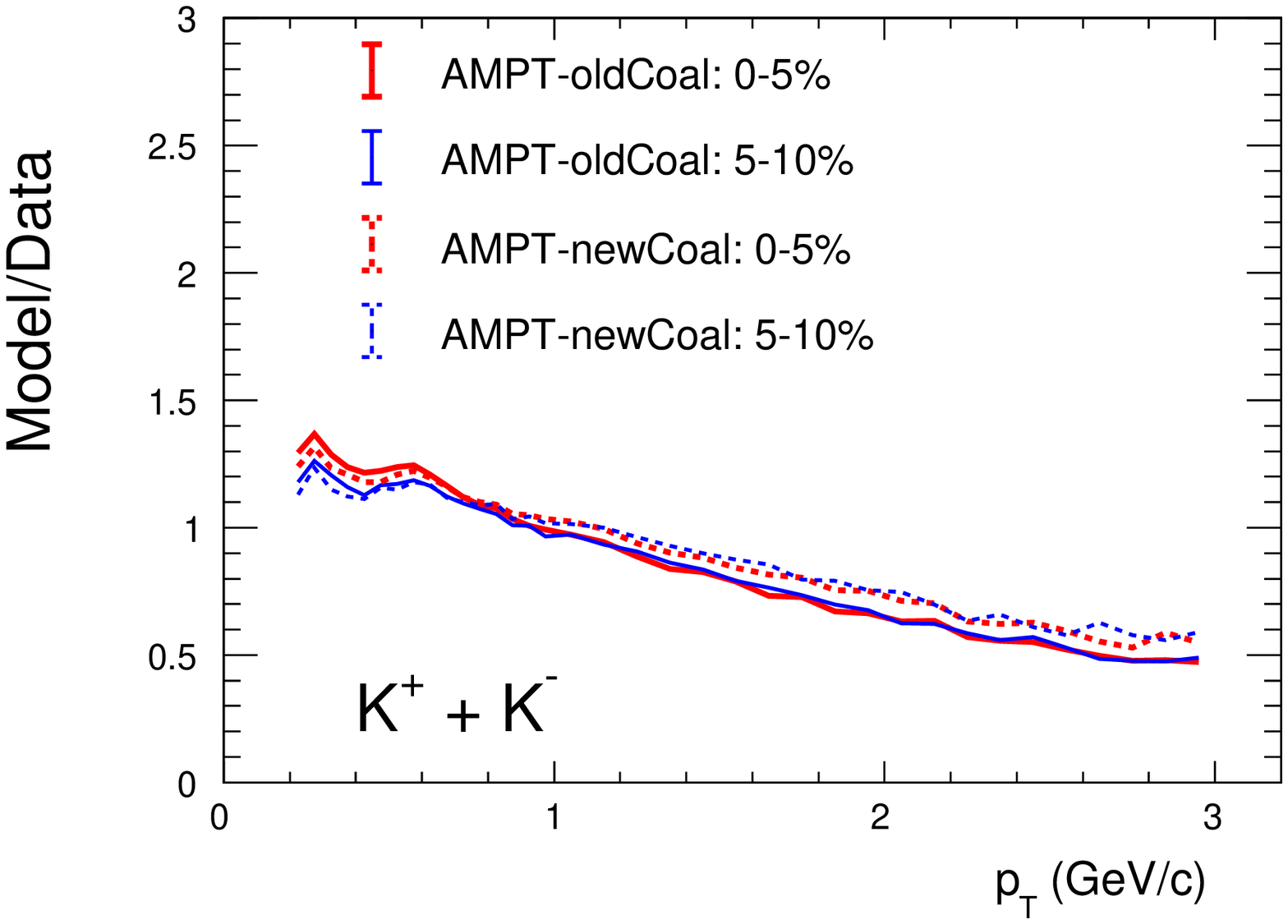}
	\includegraphics[scale=0.27]{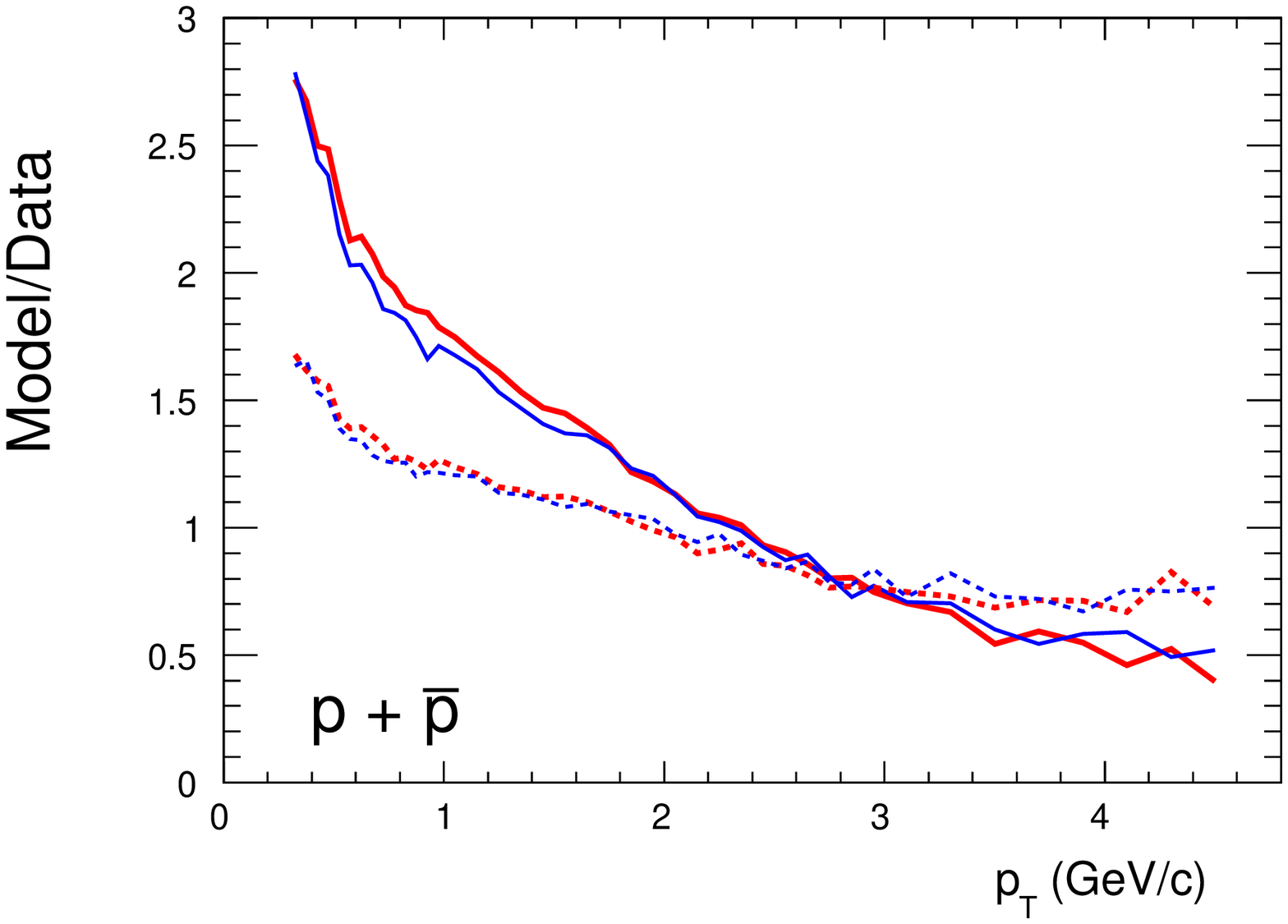}
	\caption{(Color online) The ratio of AMPT model results to the experimental data in central Pb-Pb collisions at $\mathrm{\sqrt{s_{NN}} = 2.76}$ TeV. Solid curves are results from AMPT with old quark coalescence, and dotted curves represent the new quark coalescence.}
	\label{Ratio_model_vs_data_PbPb}
\end{figure}

\begin{figure}[!htb]
	\centering
	\includegraphics[scale=0.27]{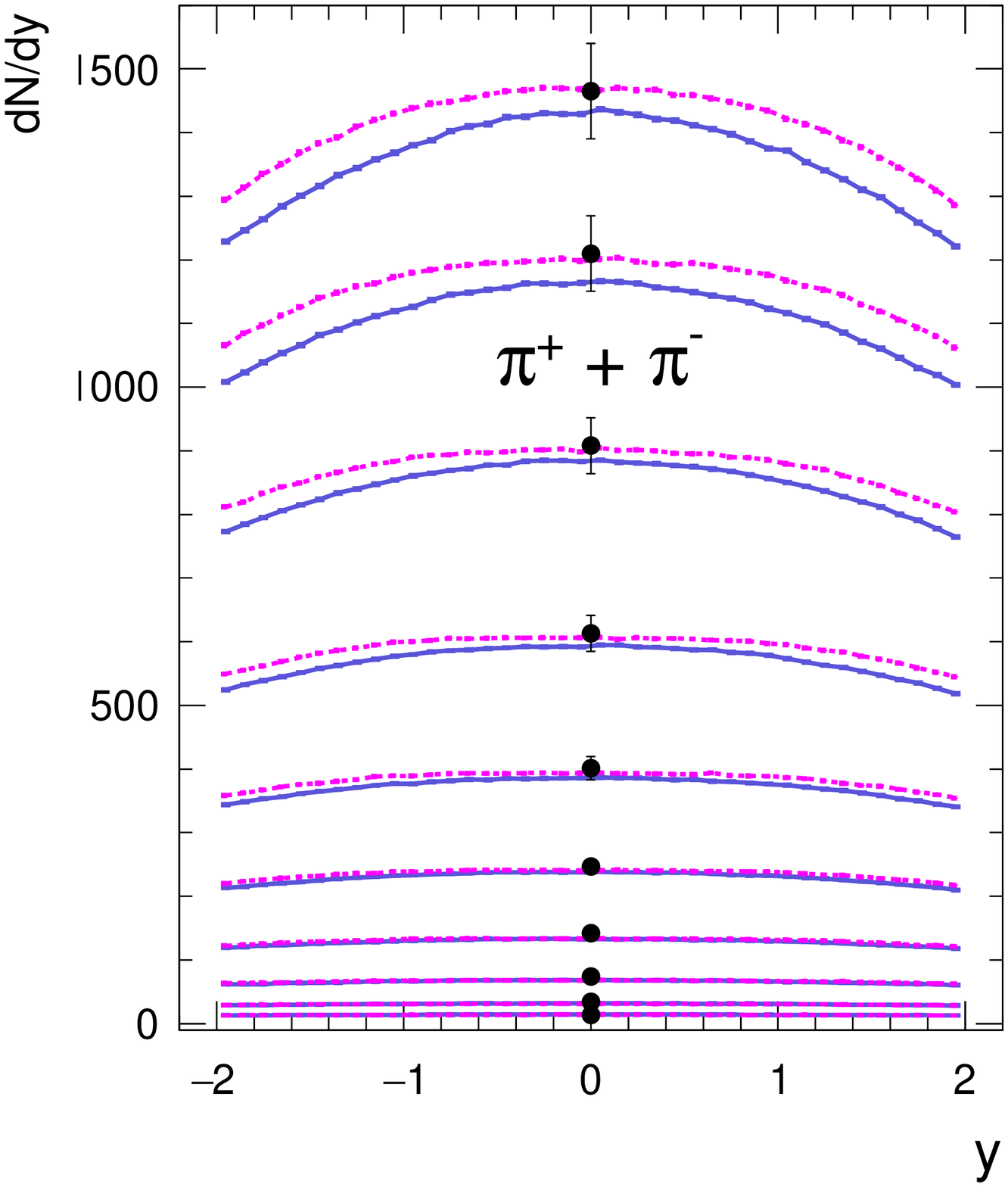}
	\includegraphics[scale=0.27]{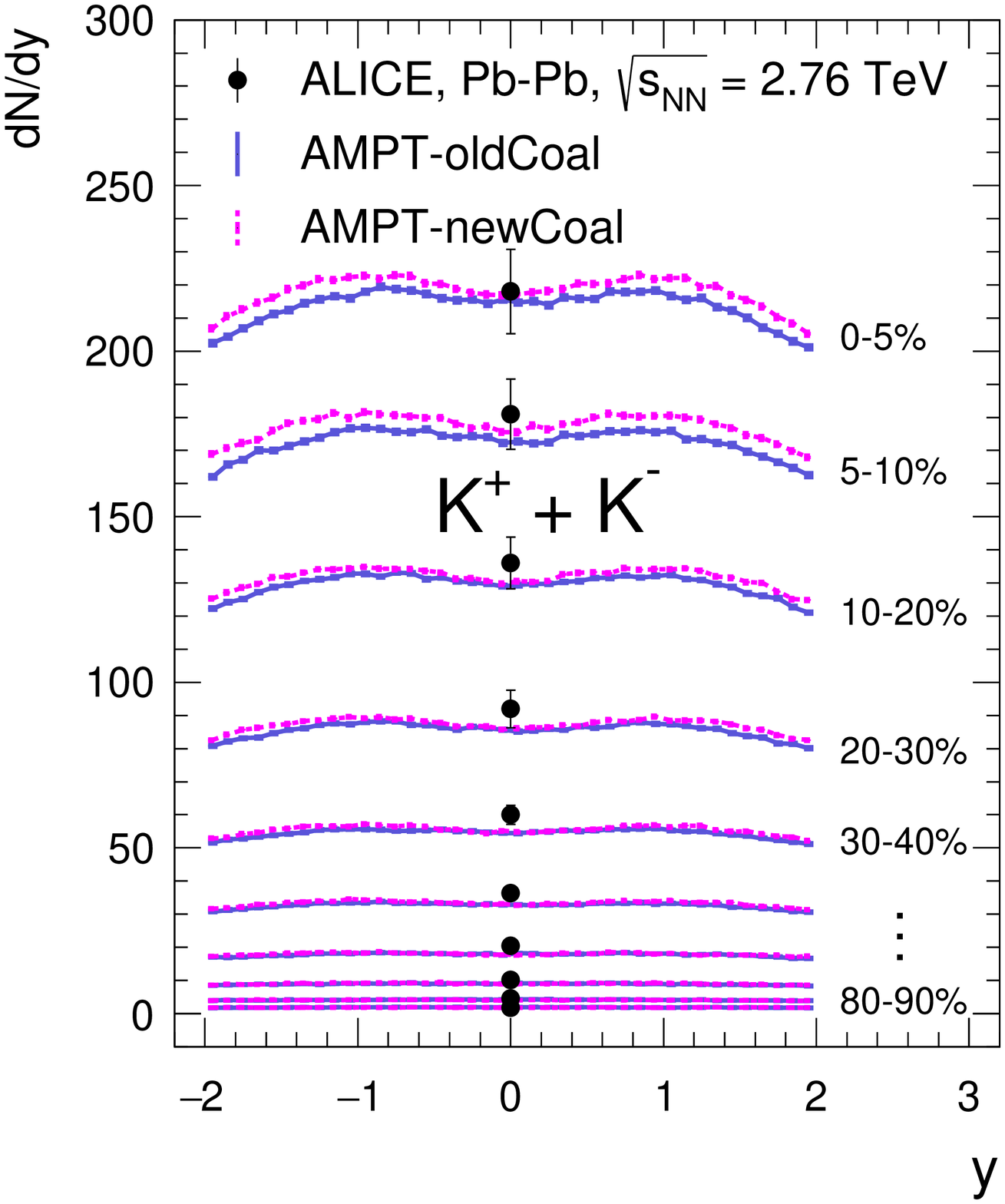}
	\includegraphics[scale=0.27]{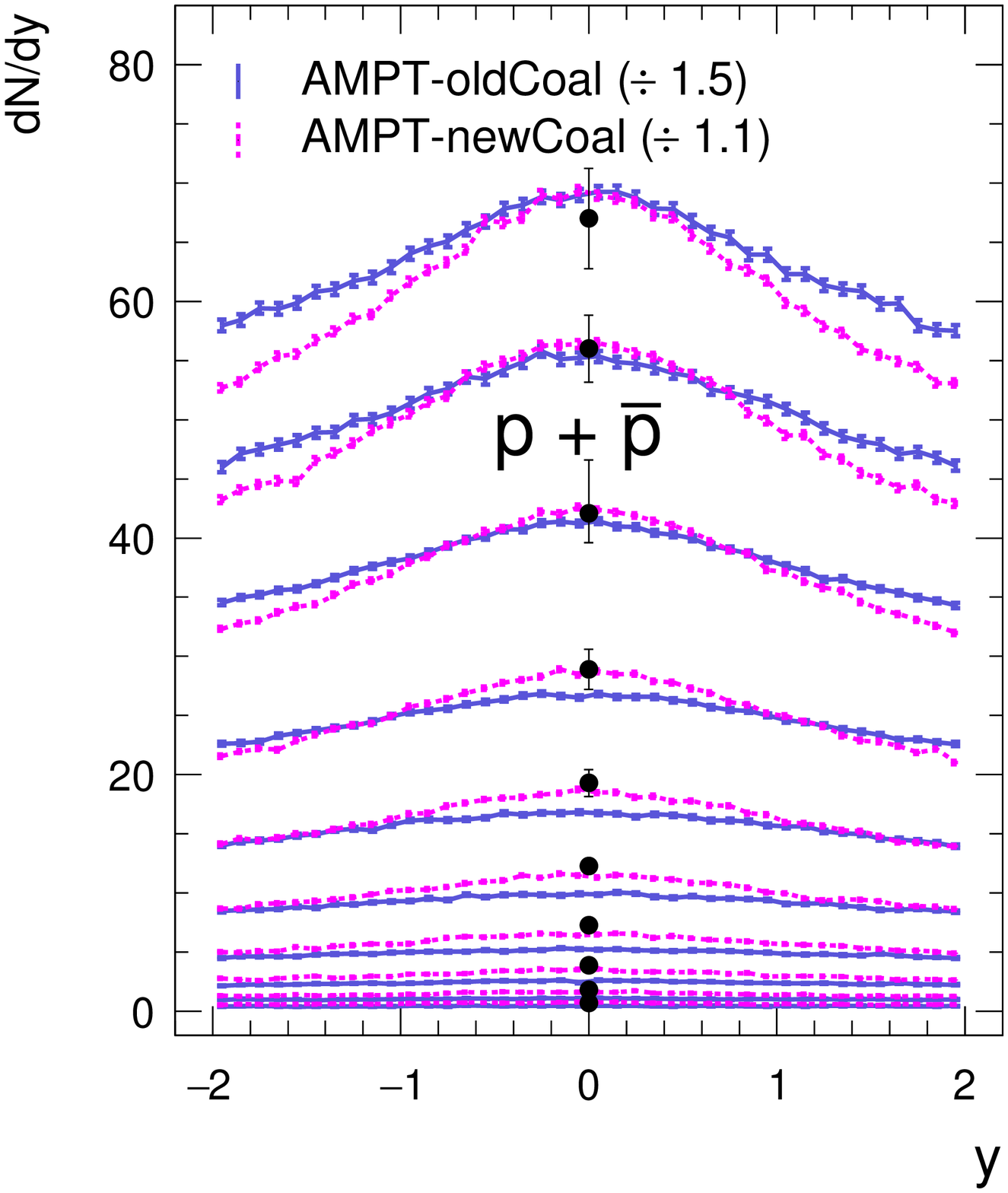}
	\caption{(Color online) $dN/dy$ distributions of $\pi^{\pm}$ (left panel), $K^{\pm}$ (middle panel) and $p$($\bar{p}$) (right panel) from the most central (0-5$\%$) to the most peripheral (80-90$\%$) Pb+Pb collisions at $\sqrt{s_{NN}}$ = 2.76 TeV. Solid curves are AMPT results  with old quark coalescence, while dotted curves represent AMPT results with new quark coalescence. Data points are experimental data~\cite{B.Abelev:2013}. Note that the proton results from AMPT model are scaled down by different factor to match the experimental data, as displayed in the figure.}
	\label{Rapidity_spectra}
\end{figure}

To check the validity of the AMPT model with new quark coalescence, we first study the $p_T$ and rapidity distributions of identified particles in the AMPT model with old and new quark coalescence. The collective properties of the hot and dense matter created in ultra-relativistic heavy-ion collisions at freeze-out stage are related to the $p_{T}$ distributions of identified particles~\cite{H.Sorge:1989,J.Chen:2008}. Figure~\ref{charged_particle_pt_spectra_pPb} shows the $p_T$ spectra of $\pi^{\pm}$, K$^{\pm}$ and $p$($\bar{p}$) in the rapidity interval of 0 $\mathrm{< y_{CMS} <}$ 0.5 in $p$-Pb collisions at $\mathrm{\sqrt{s_{NN}}}$ = 5.02 TeV from the AMPT model in comparison with the  experimental data. In general, the two AMPT model calculations can qualitatively describe the experimental data of pions and kaons in a wide $p_{T}$ range from the most central (0-5$\%$) to the most peripheral (80-100$\%$) collisions, while there are some deviations from the (anti)proton data. In order to quantify the differences, we calculate the ratios of AMPT results to the experimental data. Figure~\ref{Ratio_model_vs_data_pPb} presents the ratio results for two most central collision bins as an example. Overall, the results from the AMPT model with new quark coalescence are closer to the experimental data. The improvement is more significant on $p$($\bar{p}$), not only on the magnitude but also on the slope of the $p_T$ distributions.

Figure~\ref{charged_particle_pt_spectra_PbPb} presents the midrapidity $(|y| <$ 0.5) $p_T$ spectra of $\pi^{\pm}$, K$^{\pm}$ and $p$($\bar{p}$) in Pb-Pb collisions at $\mathrm{\sqrt{s_{NN}} = 2.76}$ TeV from the AMPT model. The ratios between the calculations and experimental data are shown in Fig.~\ref{Ratio_model_vs_data_PbPb}. Similarly to the observation from the $p$-Pb results, the AMPT model with new quark coalescence improves the description of $p$($\bar{p}$) data, while both versions describe the $\pi^{\pm}$ and K$^{\pm}$ data reasonably well. 

We then compare the charged particle rapidity density in two AMPT model calculations to the experimental data, where we only show the results for Pb-Pb collisions since the comparison for $p$-Pb collisions is similar. Figure~\ref{Rapidity_spectra} presents the $\mathrm{dN/dy}$ distributions of $\pi^{\pm}$, $K^{\pm}$ and $p$($\bar{p}$) in Pb-Pb collisions at $\mathrm{\sqrt{s_{NN}}}$ = 2.76 TeV. Qualitatively, the AMPT model calculations with new quark coalescence is in better agreement with the experimental data at midrapidity than with the old quark coalescence. Quantitatively, the overestimation of the proton data is now only 10\%, while it is up to 50\% in the original AMPT model. Our results here are consistent with the results of central collisions in Ref.~\cite{Y.C.He:2017} but with more detailed centrality intervals.

\begin{figure}[!htb]
	\centering
	\includegraphics[scale=0.32]{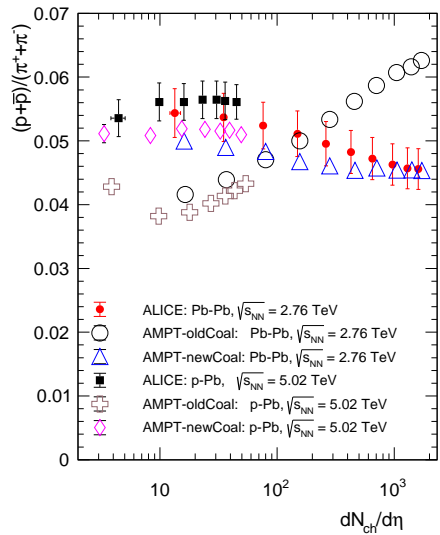}
	\caption{(Color online) The ($p$ + $\bar{p}$)/($\pi^{+}$+$\pi^{-}$) ratio as a function of $dN_{ch}$/$d\eta$ in $p$-Pb and Pb-Pb collisions at LHC energies. Open markers represent results from AMPT model calculations, while solid points are experimental data~\cite{B.Abelev:2013,B.Abelev:2014}.}
	\label{particlesum_vs_pionsum}
\end{figure}

The AMPT model with new quark coalescence also improves the description of particle ratios. Figure~\ref{particlesum_vs_pionsum} shows the ($p$ + $\bar{p}$)/($\pi^{+}$ + $\pi^{-}$) ratio within $|\eta| < 0.5$  versus the multiplicity in $p$-Pb and Pb-Pb collisions at LHC energies. The AMPT model results with new quark coalescence provide a significant improvement on the description of the experimental data compared to the results of AMPT model with old quark coalescence.  

In the following, we use the string melting version of AMPT model with new quark coalescence to study two-particle angular correlations at low $p_T$. We strictly follow the analysis process of the experiment~\cite{J.Adam:2017}, as described in detail in our early paper~\cite{L.Y.Zhang:2018}. Here we briefly introduce the analysis method. In our analysis, particles from each event are combined with particles in the same event to build the distribution of correlated pairs
\begin{equation}
	S(\Delta\eta,\Delta\phi) = \frac{\mathrm{d^2N_{pairs}^{signal}}}{d{\Delta\eta}d\Delta\phi},  
\end{equation}
while they are combined with particles from other events to build the reference distribution
\begin{equation}
 B(\Delta\eta,\Delta\phi) = \frac{\mathrm{d^2N_{pairs}^{mixed}}}{d{\Delta\eta}d\Delta\phi}.
\end{equation}  
Each event is mixed with 10 other events in this study to improve the statistical power of the reference estimation, and the impact parameter direction of the collisions in AMPT events is rotated randomly in the transverse plane for the $B$($\Delta\eta$,$\Delta\phi$) calculations. Another check by mixing event with similar event plane direction has been carried out, and the difference between different background reconstructions is negligible. The two-particle correlation function is then built as 
\begin{equation}
	C(\Delta\eta,\Delta\phi) = \frac{S(\Delta\eta,\Delta\phi)/\mathrm{N_{pairs}^{signal}}}{B(\Delta\eta,\Delta\phi)/\mathrm{N_{pairs}^{mixed}}}.
\end{equation}
A one-dimensional $\Delta\phi$ correlation function can be constructed 
from the $C$($\Delta\eta,\Delta\phi$) by integrating over $\Delta\eta$ as
\begin{equation}
	C(\Delta\phi) = A\times\frac{\int S(\Delta\eta,\Delta\phi)d\Delta\eta}{\int B(\Delta\eta,\Delta\phi)d\Delta\eta},
\end{equation}
where the normalization constant $A$ is given by $\mathrm{N_{pairs}^{mixed}/N_{pairs}^{signal}}$.

\begin{figure}[!htb]
	\centering
	\includegraphics[scale=0.25]{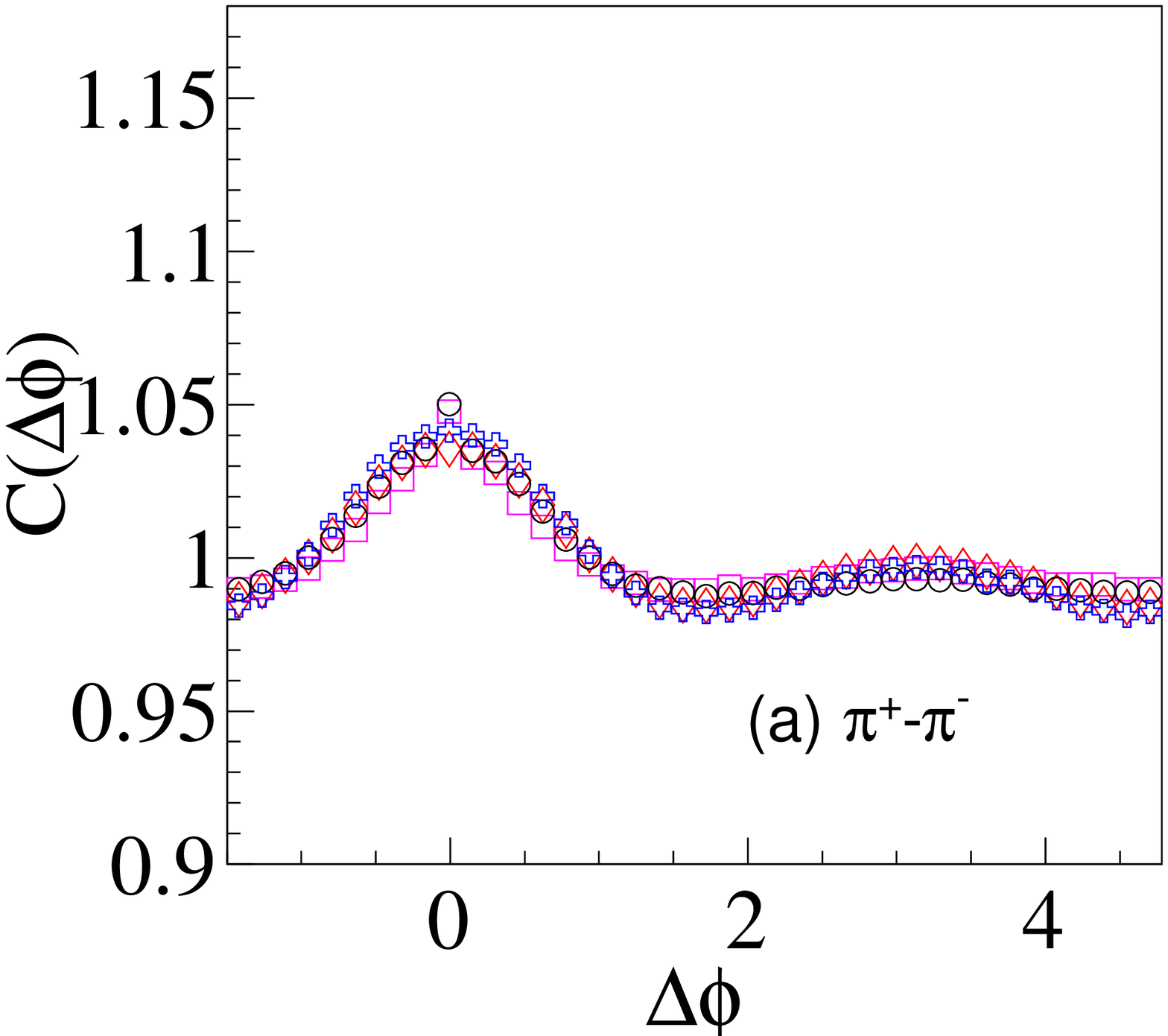}
	\includegraphics[scale=0.25]{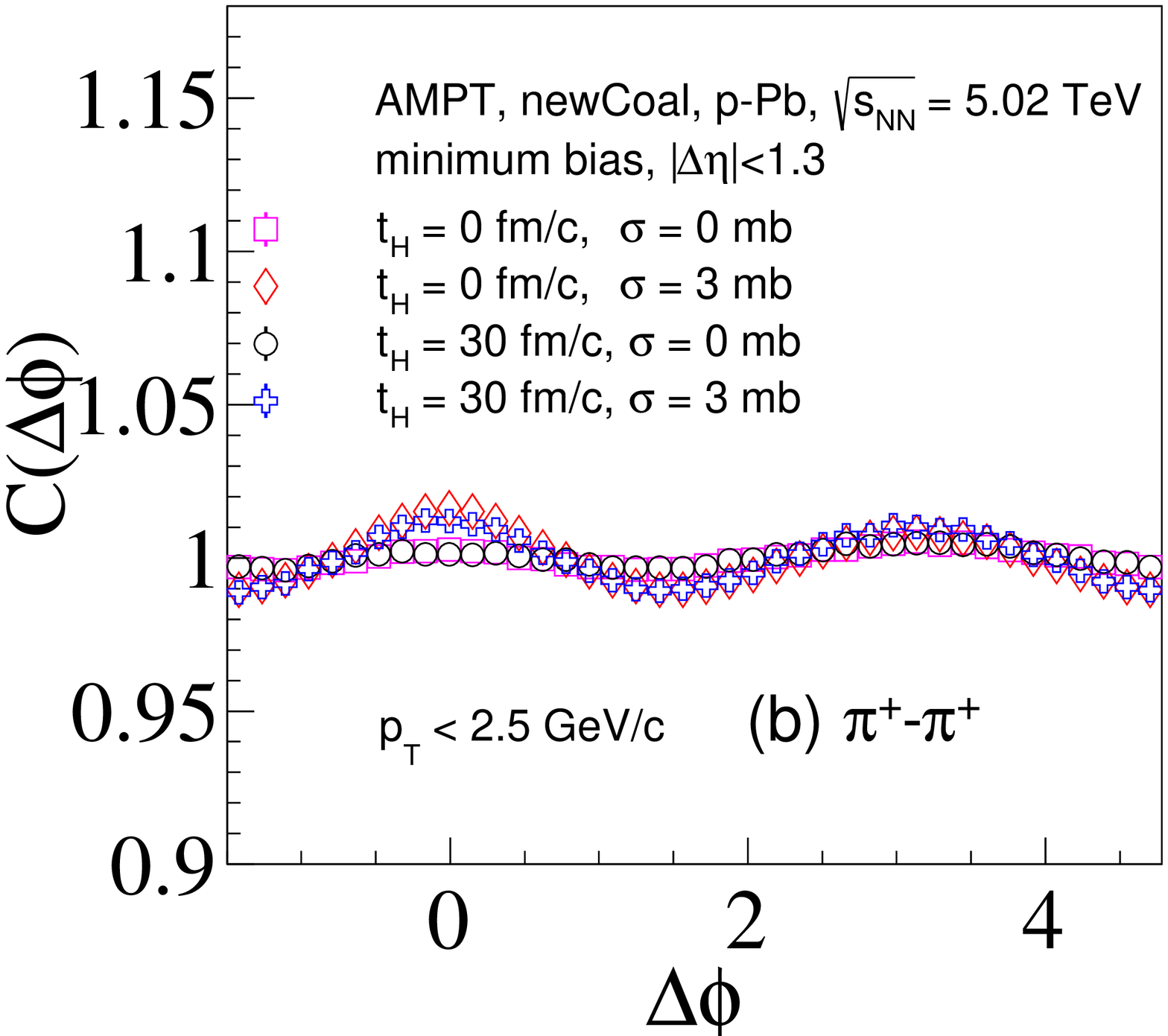}
	\includegraphics[scale=0.25]{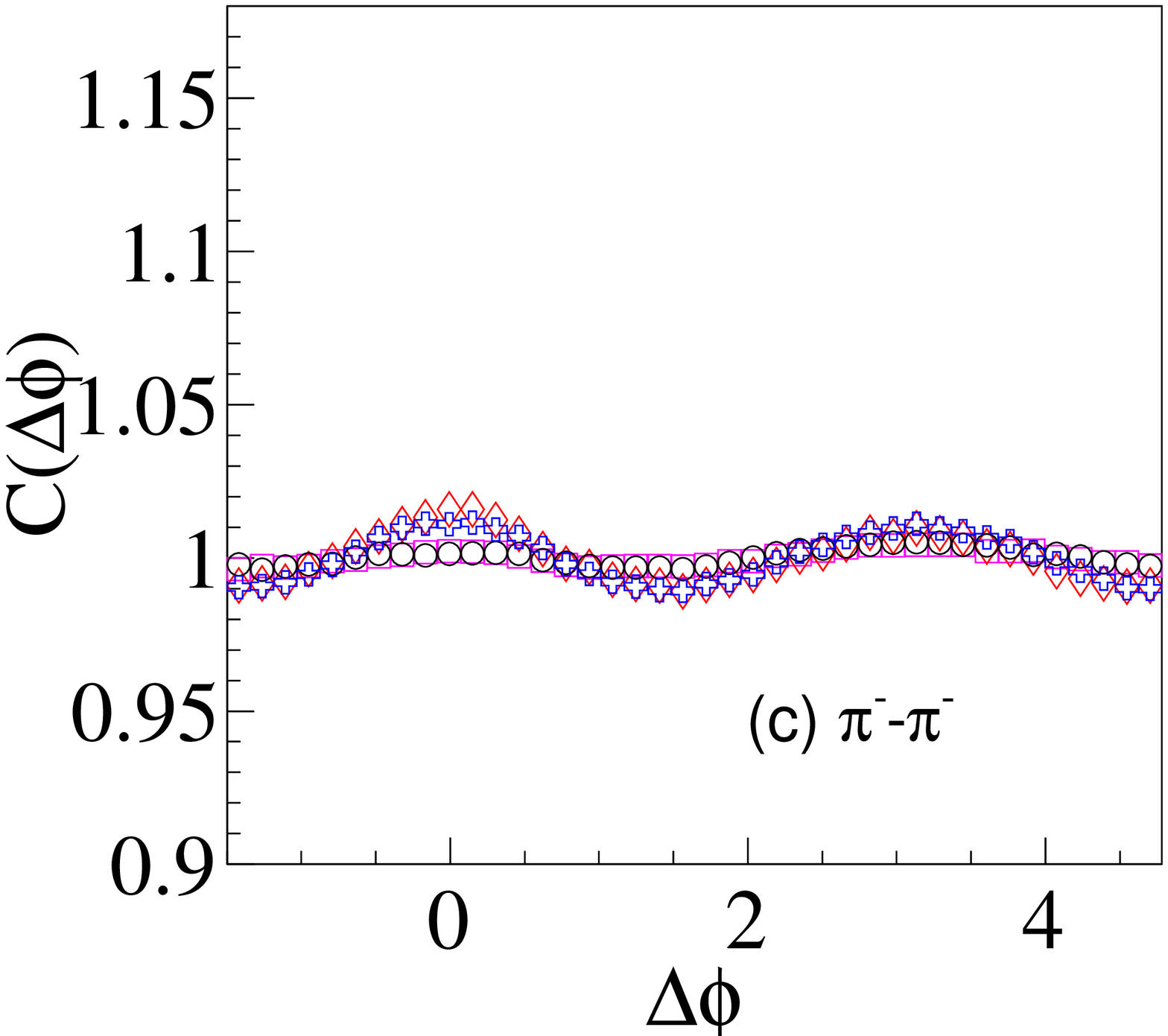}
	\caption{(Color online) One-dimensional $\Delta\phi$ correlation functions for $\pi^{+}$-$\pi^{-}$, $\pi^{+}$-$\pi^{+}$, and $\pi^{-}$-$\pi^{-}$ in $p$-Pb collisions at $\mathrm{\sqrt{s_{NN}}= 5.02}$ TeV from the AMPT model with $\mathrm{t_{H}}$ = 0 fm/c (hadronic scattering turned off) or $\mathrm{t_{H}}$ = 30 fm/c (hadronic scattering turned on) and parton cross section $\sigma$ = 0 mb or 3 mb.}  
	\label{pion_pair_correlation_pPb}
\end{figure}

\begin{figure}[!htb]
	\centering
	\includegraphics[scale=0.25]{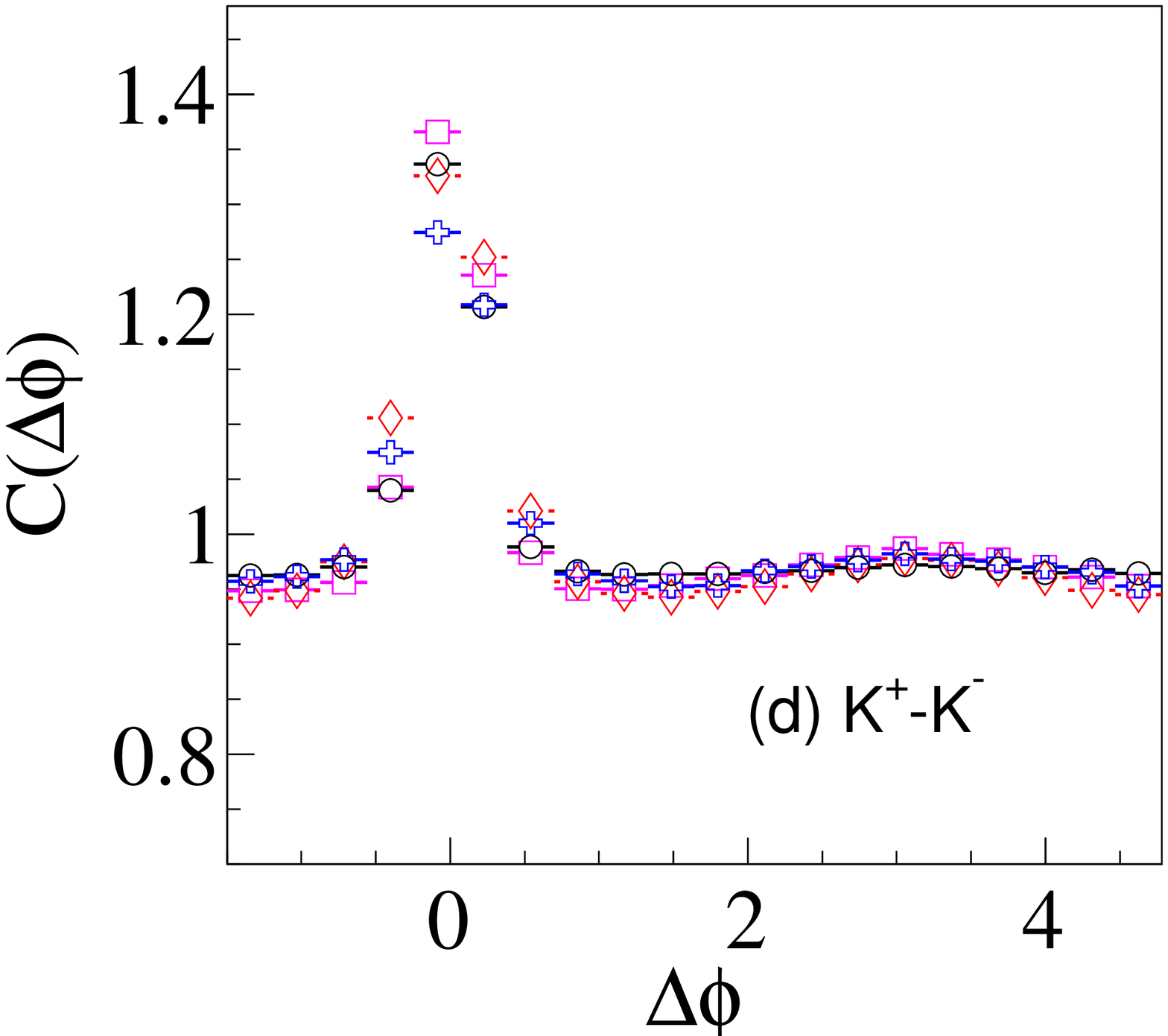}
	\includegraphics[scale=0.25]{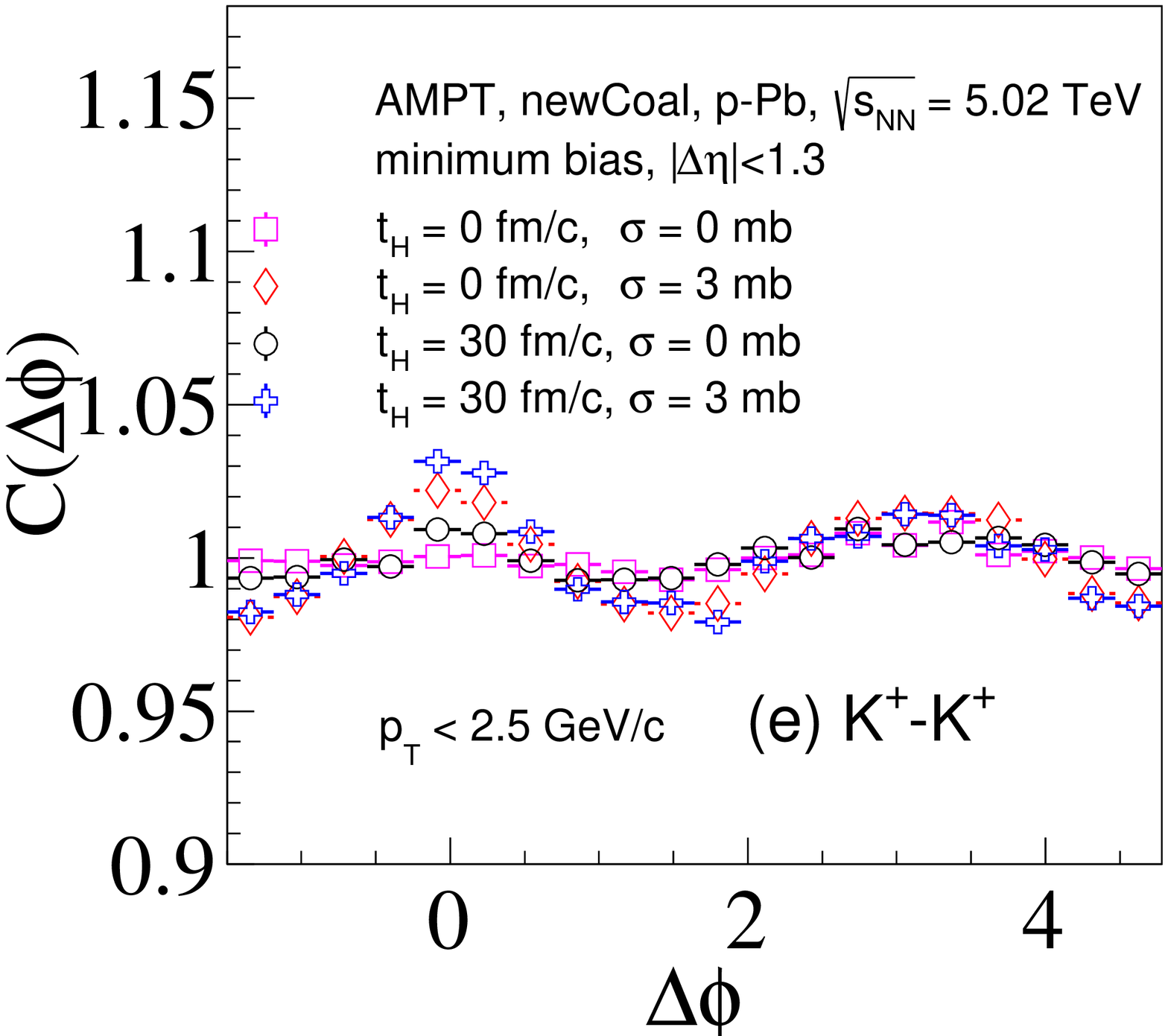}
	\includegraphics[scale=0.25]{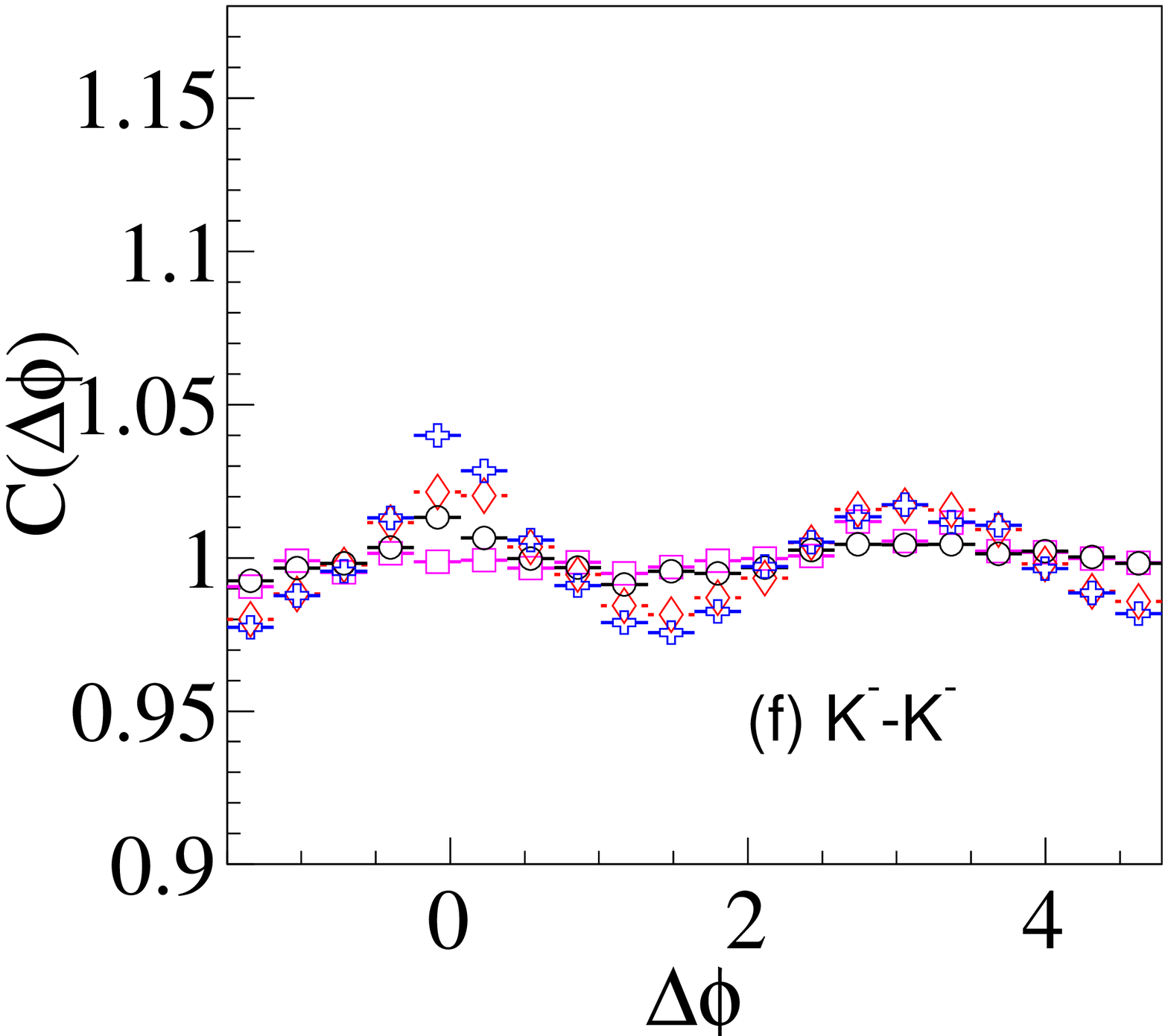}
	\caption{(Color online) One-dimensional $\Delta\phi$ correlation functions for K$^{+}$-K$^{-}$, K$^{+}$-K$^{+}$, and K$^{-}$-K$^{-}$ in $p$-Pb collisions at $\mathrm{\sqrt{s_{NN}}= 5.02}$ TeV from the AMPT model with $\mathrm{t_{H}}$ = 0 fm/c (hadronic scattering turned off) or $\mathrm{t_{H}}$ = 30 fm/c (hadronic scattering turned on) and parton cross section $\sigma$ = 0 mb or 3 mb.}
	\label{kaon_pair_correlation_pPb}
\end{figure}

\begin{figure}[!htb]
	\centering
	\includegraphics[scale=0.25]{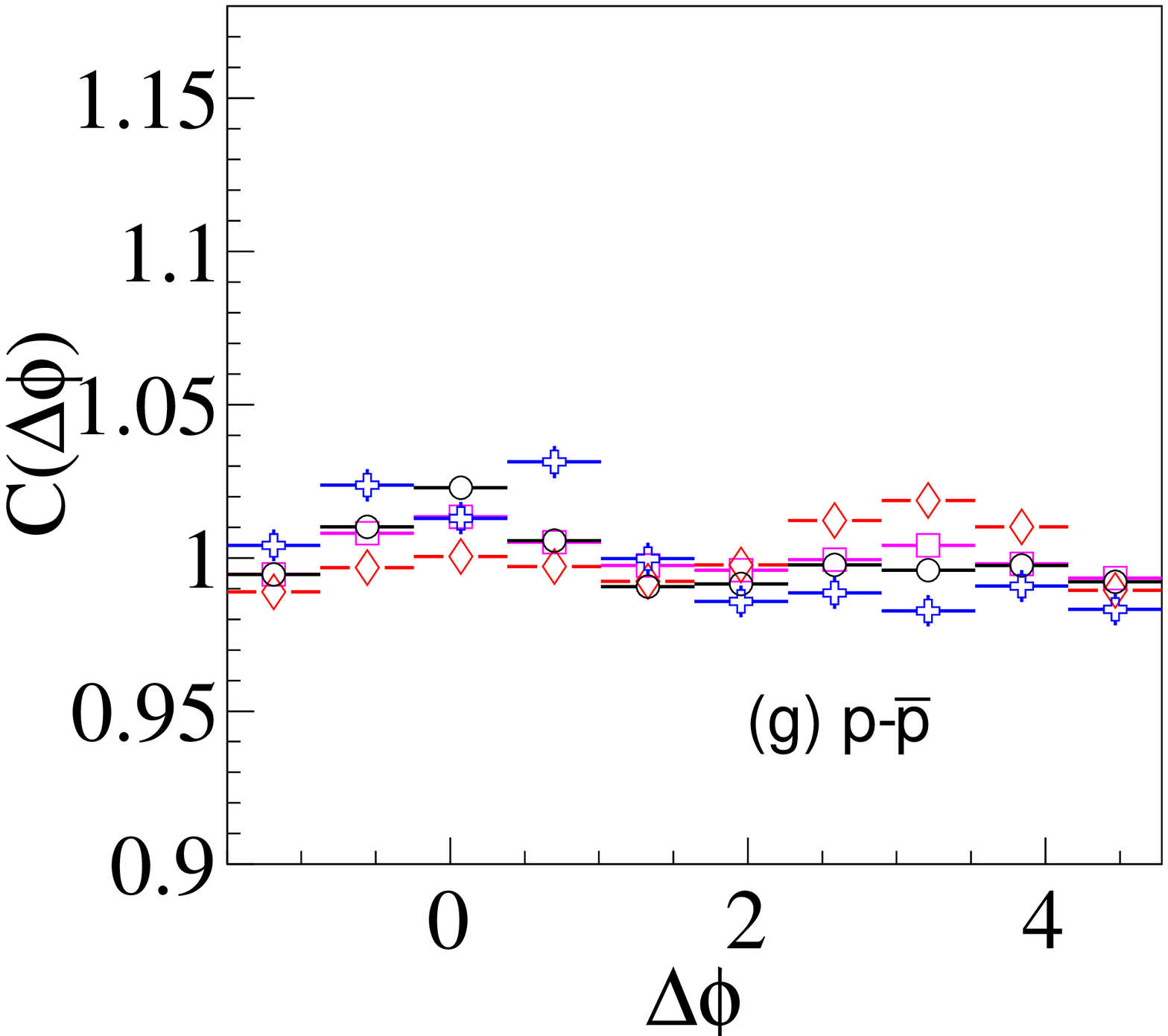}
	\includegraphics[scale=0.25]{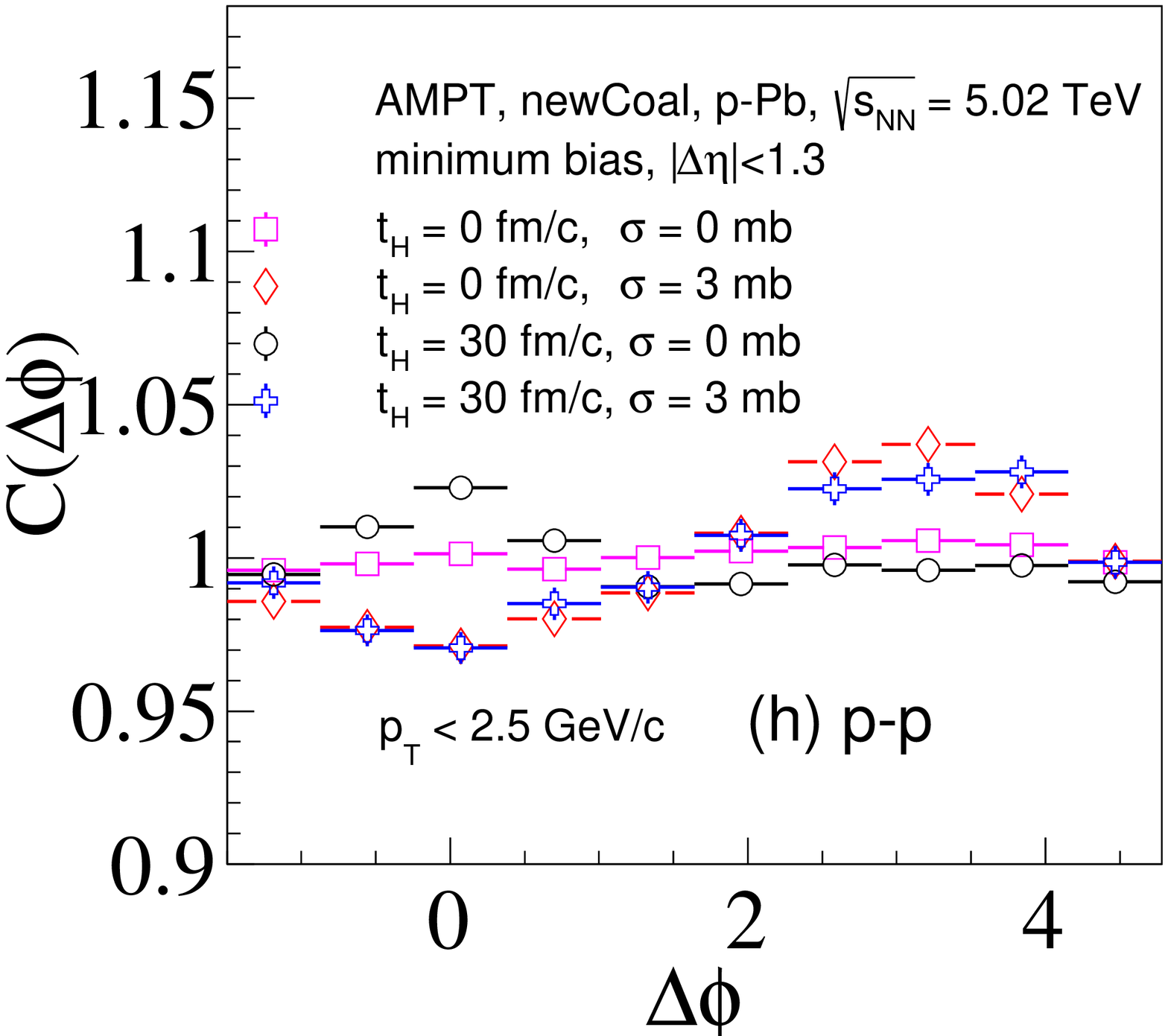}
	\includegraphics[scale=0.25]{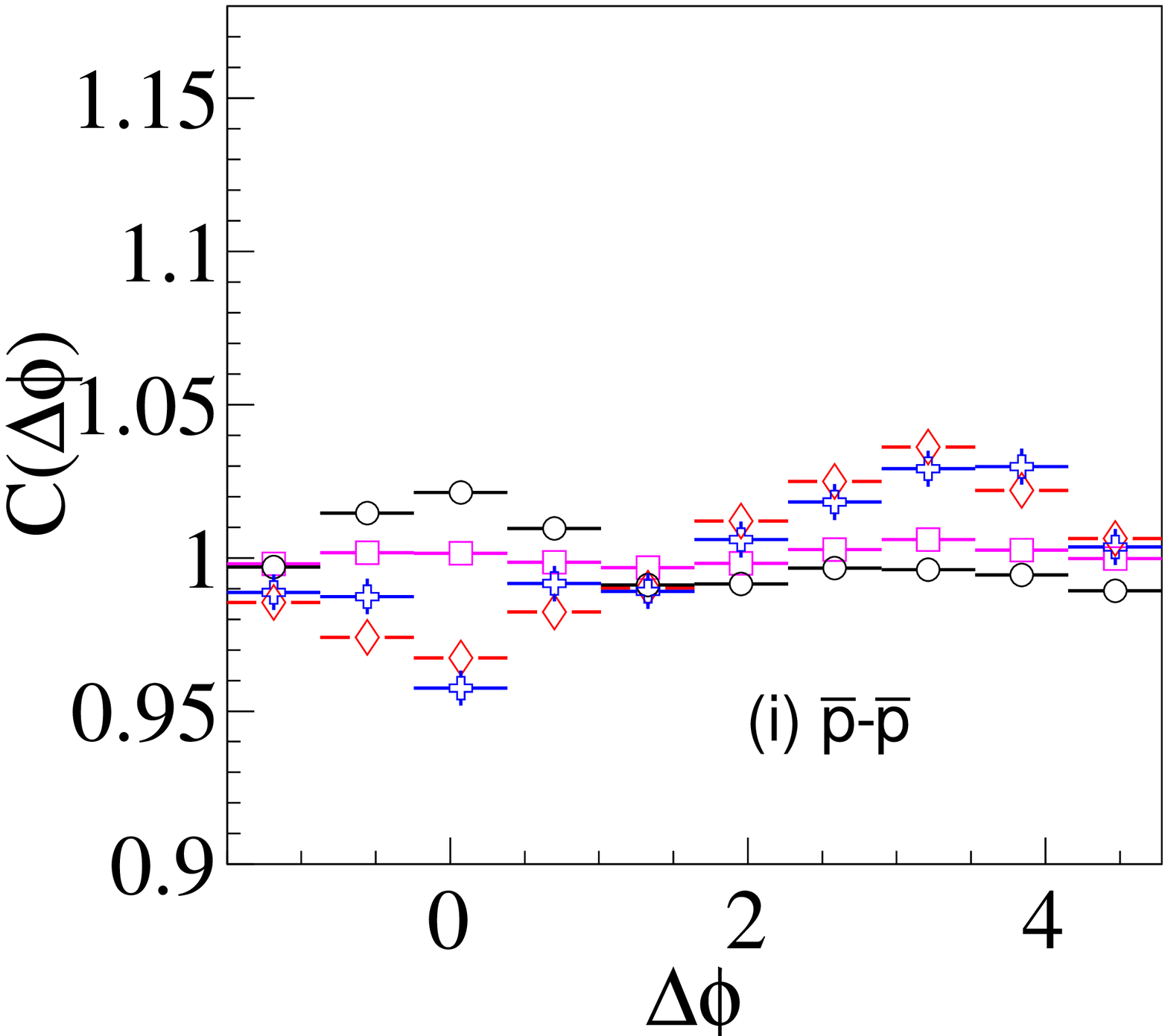}
	\caption{(Color online) One-dimensional $\Delta\phi$ correlation functions for $p$-$\bar{p}$, $p$-$p$, and $\bar{p}$-$\bar{p}$ in $p$-Pb collisions at $\mathrm{\sqrt{s_{NN}}= 5.02}$ TeV from the AMPT model with $\mathrm{t_{H}}$ = 0 fm/c (hadronic scattering turned off) or $\mathrm{t_{H}}$ = 30 fm/c (hadronic scattering turned on) and parton cross section $\sigma$ = 0 mb or 3 mb.}
	\label{proton_pair_correlation_pPb}
\end{figure}

Figures~\ref{pion_pair_correlation_pPb},~\ref{kaon_pair_correlation_pPb} and~\ref{proton_pair_correlation_pPb} show the particle-antiparticle and particle-particle pair correlation functions along the $|\Delta\phi|$ axis from the AMPT model
for pions, kaons, and (anti)protons respectively. The hadronic cascade termination time $\mathrm{t_{H}}$ = 0 or 30 fm/c has been chosen to investigate the hadronic rescattering contributions to the correlation functions, with $\mathrm{t_{H}}$ = 0 fm/c representing the case with the hadronic cascade turned off. 
In addition, the parton cross section $\sigma$ = 0 mb or 3 mb is chosen to study the parton rescattering contributions to the correlation functions, with $\sigma$ = 0 mb representing the case with no parton interactions. 


The meson pair correlation functions show a near side peak structure and a lower amplitude peak in the away side, indicating that minijets and resonance decays in the AMPT model give a dominant contribution to the structure of these two-particle angular correlations in $p$-Pb collisions, similar to that observed in $p$-$p$ collisions~\cite{L.Y.Zhang:2018}. For the same-charge $\pi$-$\pi$ and K-K correlation functions from the AMPT model, the correlation functions with a sufficient hadronic cascade termination time $\mathrm{t_{H}}$ = 30 fm/c is almost the same as the results without hadronic interactions ($\mathrm{t_{H}}$ = 0 fm/c) in near side and away side. However, the correlation functions with parton cascade show higher magnitudes for the near-side and away-side peaks in comparison to that without parton cascade ($\sigma$ = 0 mb). The results suggest that the parton scatterings are much more important than hadronic scatterings for the same-charge correlation functions. 

Interestingly, we observe in Fig.~\ref{proton_pair_correlation_pPb} a pronounced depression on near side same-charge baryon-baryon and antibaryon-antibaryon (but not baryon-antibaryon) correlation functions. The results include a clear depression structure on the near side and a strong enhancement on the away side, in contrast to same-charge meson correlations. 
The scenario with $\sigma$ = 0 mb and $\mathrm{t_{H}}$ = 30 fm/c is similar to the default AMPT version, which is not expected to show a depression in baryon pair angular correlations~\cite{L.Y.Zhang:2018}. By comparing results with different parton cross sections and taking into account the fact that the string melting version of AMPT with old quark coalescence does not show anti-correlations~\cite{L.Y.Zhang:2018}, it is clear that both parton cascade and the new quark coalescence are important to describe the depression of baryon-baryon and antibaryon-antibaryon correlations.

\begin{figure}[!htb]
	\centering
	\includegraphics[scale=0.25]{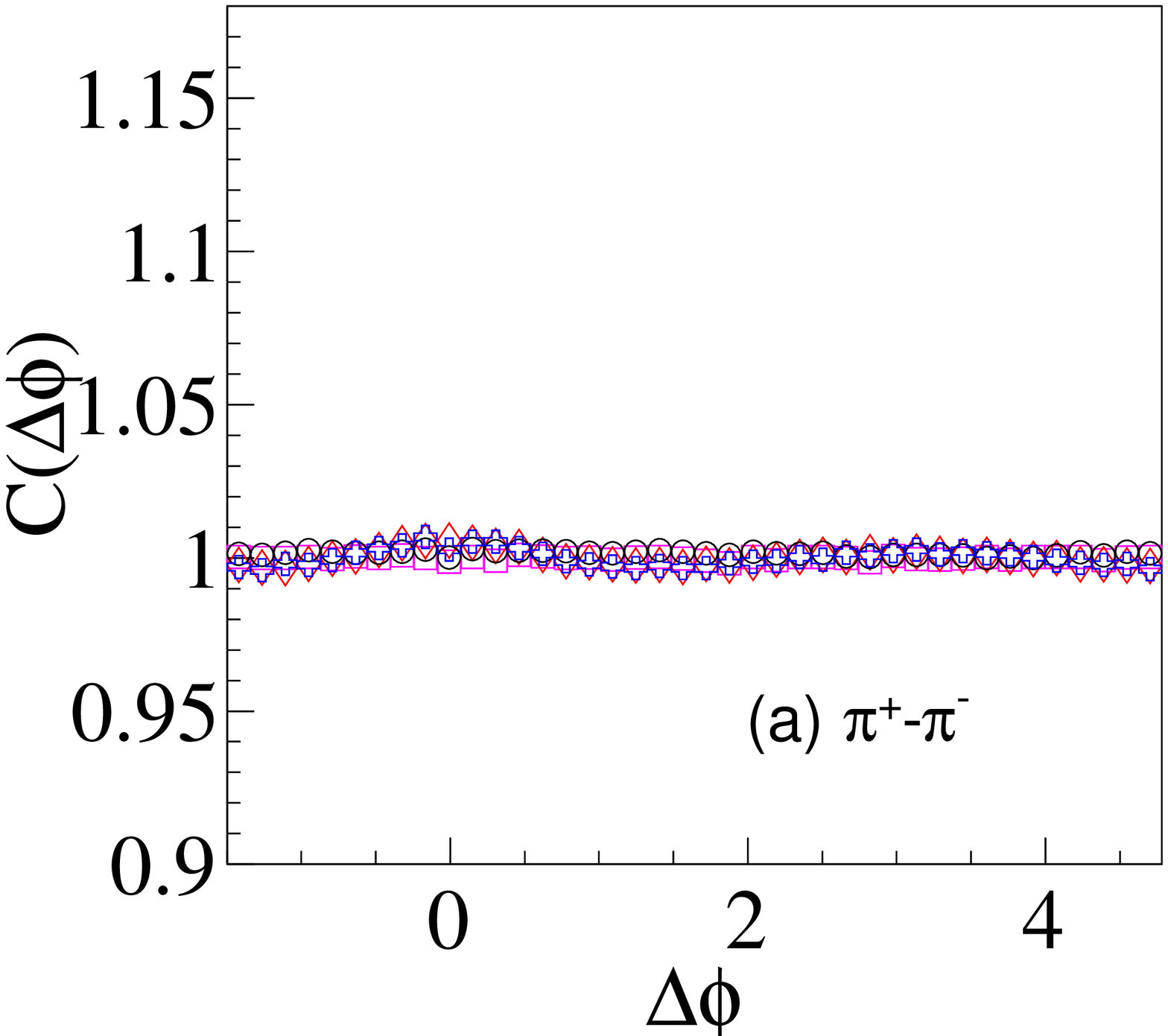}
	\includegraphics[scale=0.25]{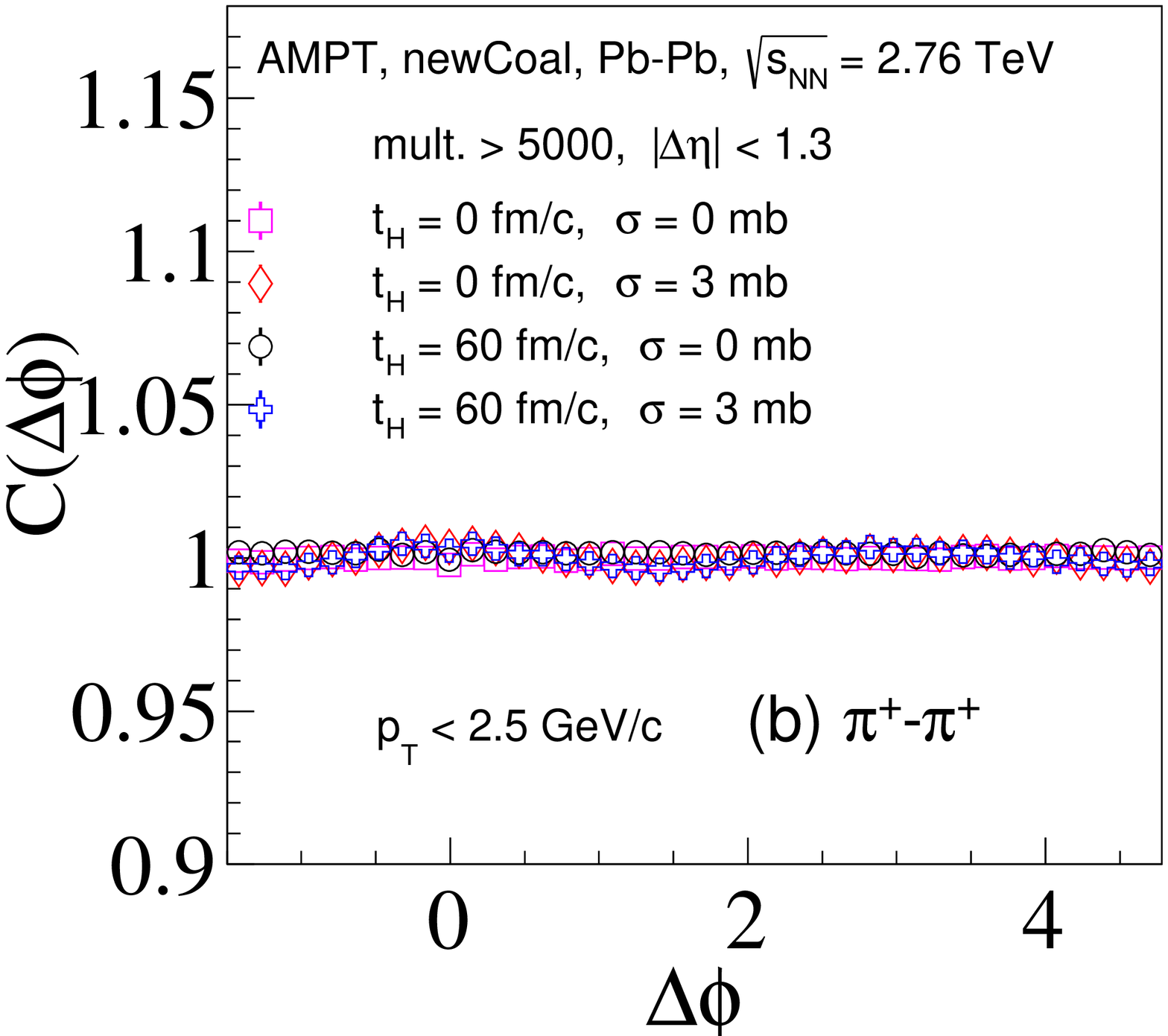}
	\includegraphics[scale=0.25]{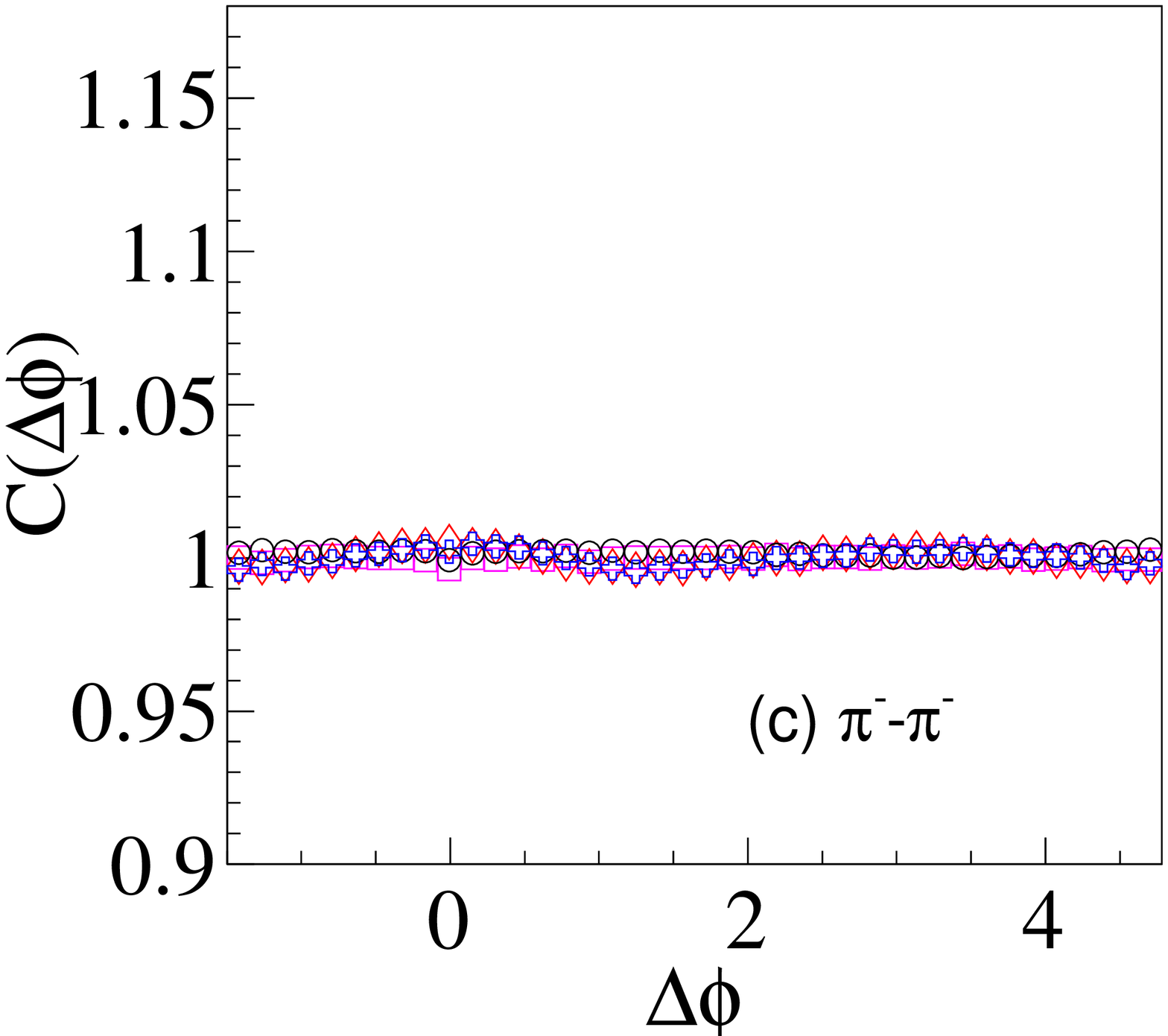}
	\caption{(Color online) One-dimensional $\Delta\phi$ correlation functions for $\pi^{+}$-$\pi^{-}$, $\pi^{+}$-$\pi^{+}$, and $\pi^{-}$-$\pi^{-}$ in Pb-Pb collisions $\mathrm{\sqrt{s_{NN}}= 2.76}$ TeV from the AMPT model with $\mathrm{t_{H}}$ = 0 fm/c (hadronic scattering turned off) or $\mathrm{t_{H}}$ = 60 fm/c (hadronic scattering turned on) and parton cross section $\sigma$ = 0 mb or 3 mb.}
	\label{pion_pair_correlation_PbPb}
\end{figure}
\begin{figure}[!htb]
	\centering
	\includegraphics[scale=0.25]{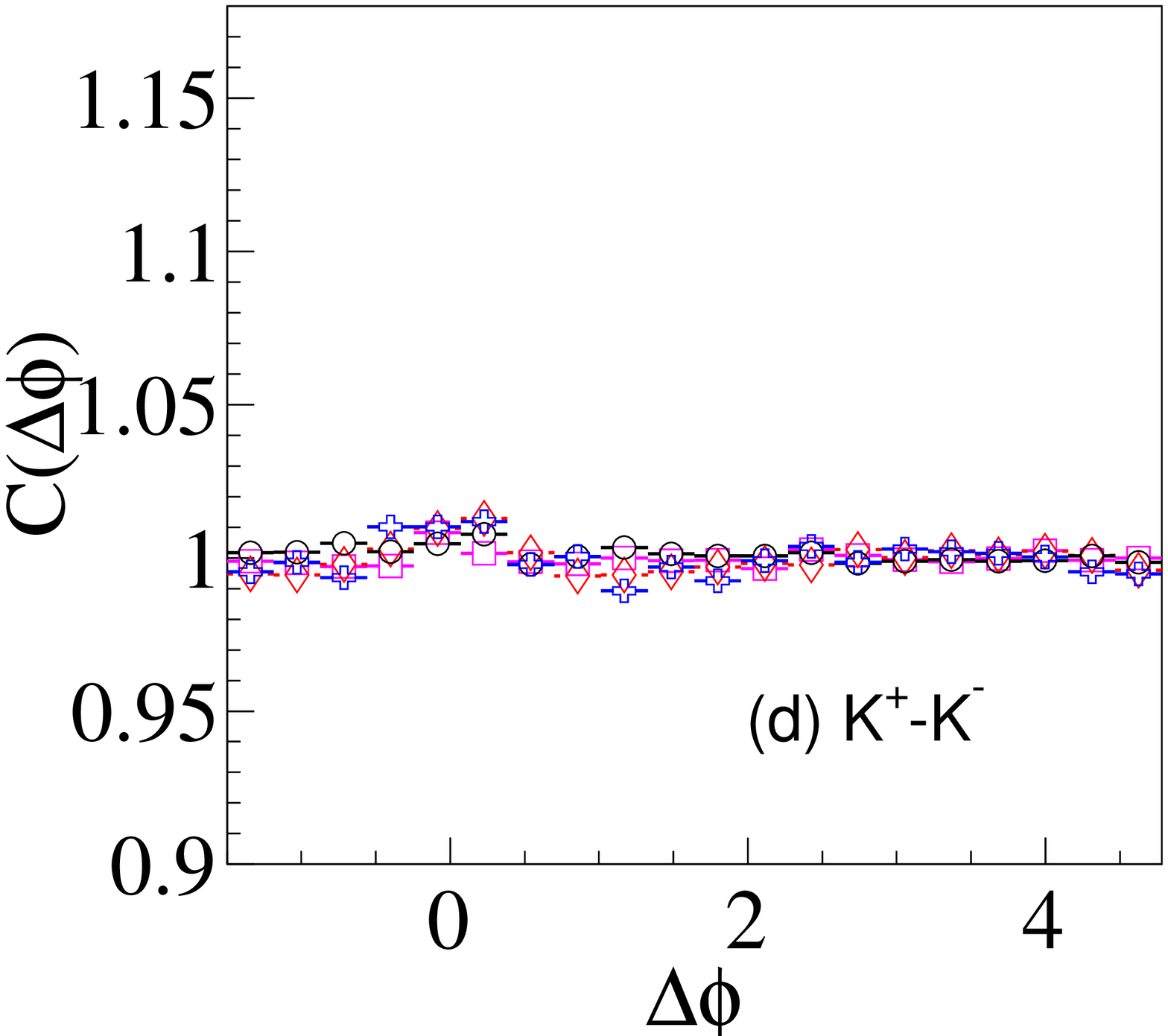}
	\includegraphics[scale=0.25]{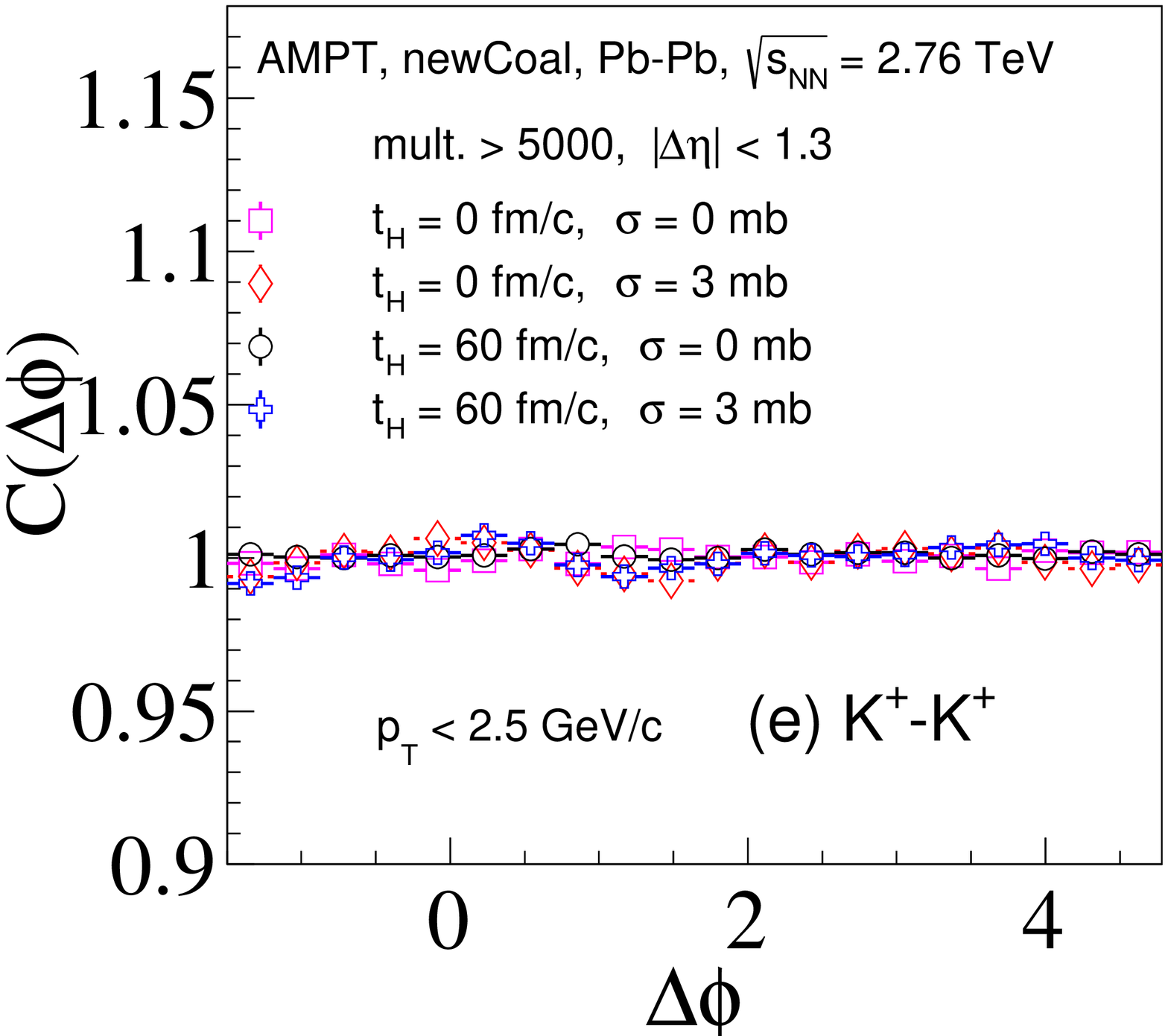}
	\includegraphics[scale=0.25]{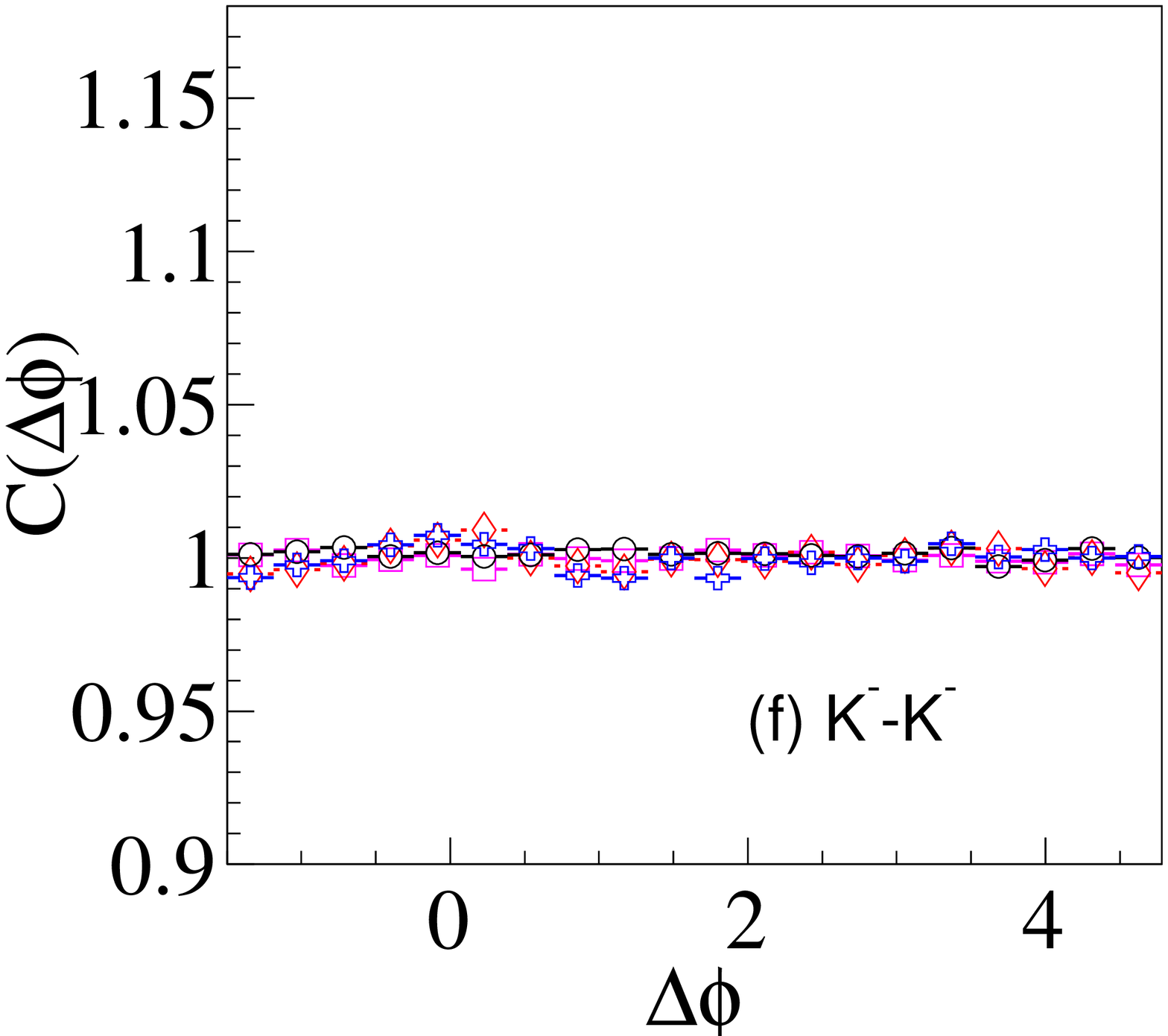}
	\caption{(Color online) One-dimensional $\Delta\phi$ correlation functions for K$^{+}$-K$^{-}$, K$^{+}$-K$^{+}$ and K$^{-}$-K$^{-}$ in Pb-Pb collisions $\mathrm{\sqrt{s_{NN}}= 2.76}$ TeV from the AMPT model with $\mathrm{t_{H}}$ = 0 fm/c (hadronic scattering turned off) or $\mathrm{t_{H}}$ = 60 fm/c (hadronic scattering turned on) and parton cross section $\sigma$ = 0 mb or 3 mb.}
	\label{kaon_pair_correlation_PbPb}
\end{figure}
\begin{figure}[!htb]
	\centering
	\includegraphics[scale=0.25]{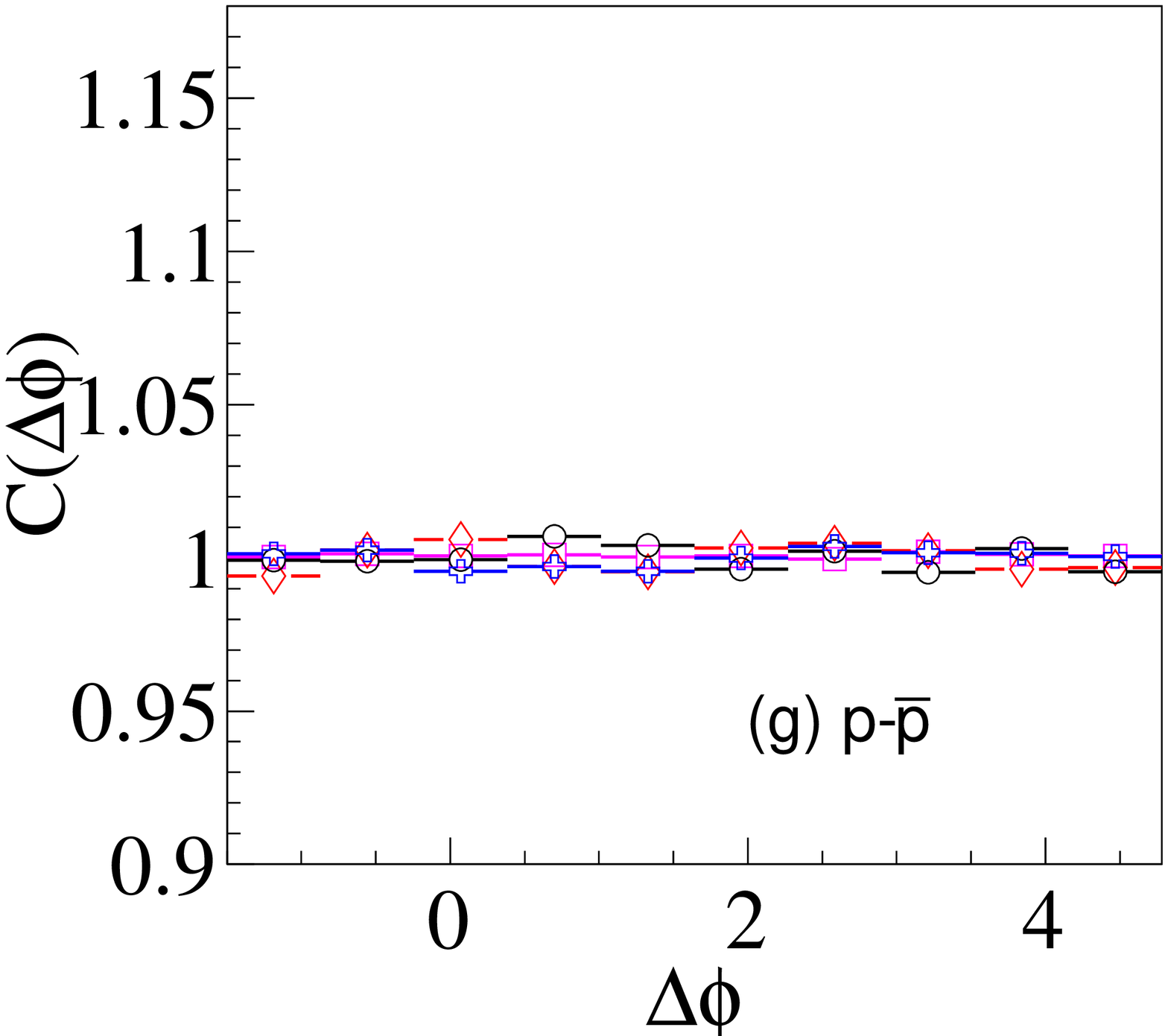}
	\includegraphics[scale=0.25]{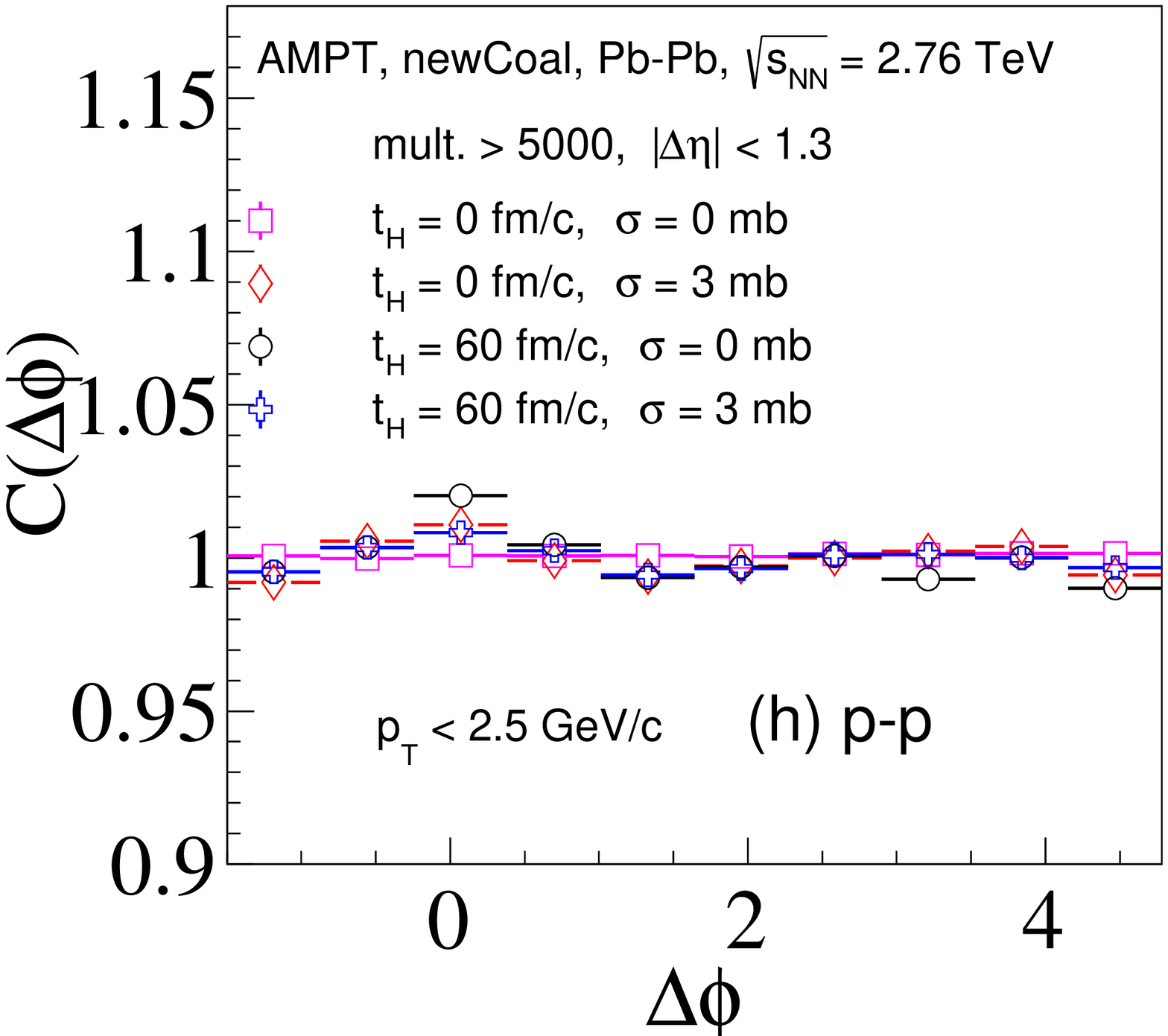}
	\includegraphics[scale=0.25]{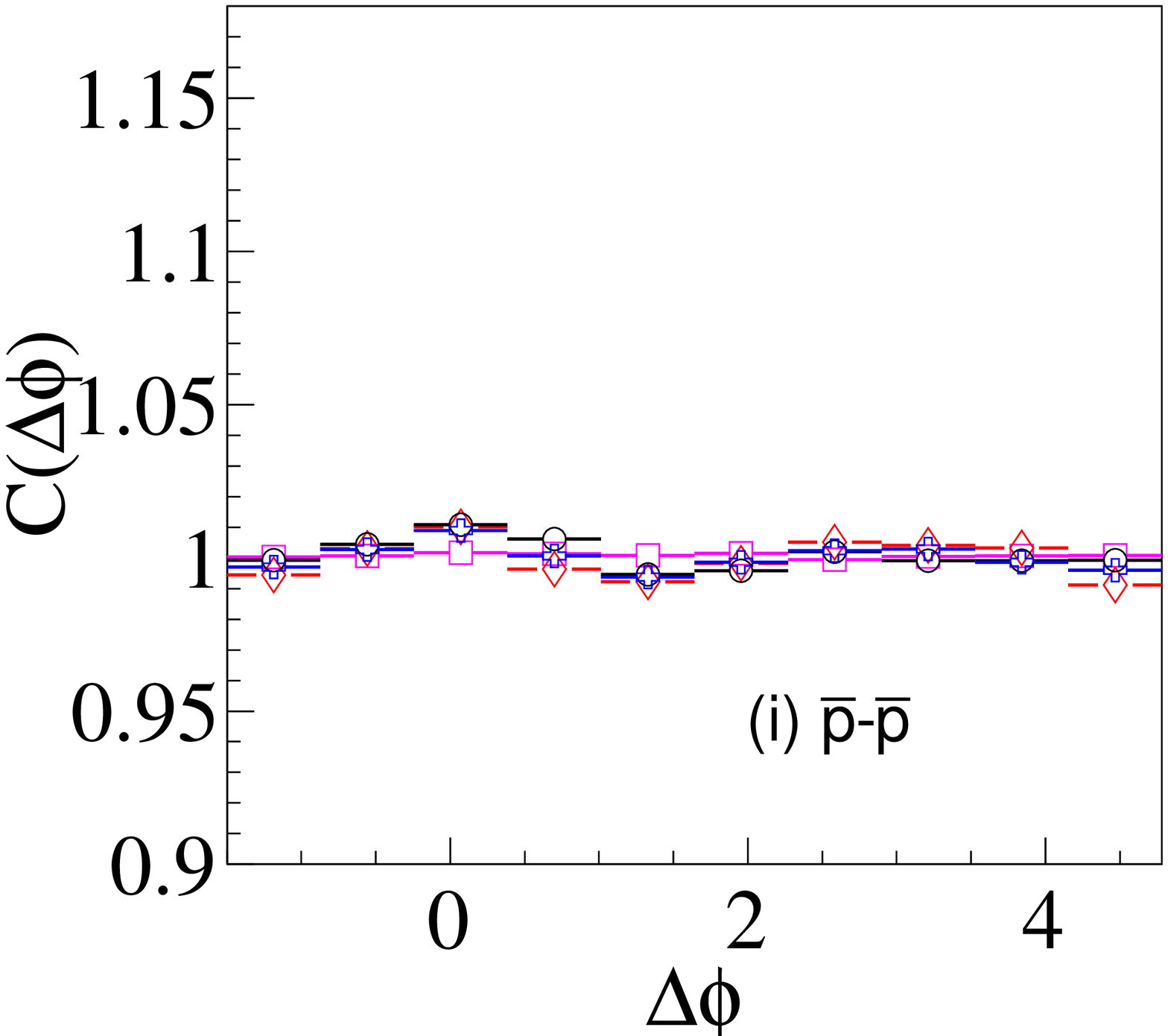}
	\caption{(Color online) One-dimensional $\Delta\phi$ correlation functions for $p$-$\bar{p}$, $p$-$p$ and $\bar{p}$-$\bar{p}$ in Pb-Pb collisions $\mathrm{\sqrt{s_{NN}}= 2.76}$ TeV from the AMPT model with $\mathrm{t_{H}}$ = 0 fm/c (hadronic scattering turned off) or $\mathrm{t_{H}}$ = 60 fm/c (hadronic scattering turned on) and parton cross section $\sigma$ = 0 mb or 3 mb.}
	\label{proton_pair_correlation_PbPb}
\end{figure}

Results for central and semi-central Pb-Pb collisions at $\mathrm{\sqrt{s_{NN}}= 2.76}$ TeV 
on the angular correlation functions of $\pi^{\pm}$, $K^{\pm}$, and $p$($\bar{p}$) pairs are shown in Figs.~\ref{pion_pair_correlation_PbPb},~\ref{kaon_pair_correlation_PbPb} and Fig.~\ref{proton_pair_correlation_PbPb}, respectively. A larger $\mathrm{t_H}$ parameter than the one used in $pp$ and $p$-Pb collisions is chosen to be safe because a longer hadronic scattering period may be expected. We first study central and semi-central events with multiplicity larger than 5000. Here, multiplicity is defined as the number of charged tracks within $|y| <$ 0.8 and $p_{T}$ $<$ 2.5 GeV/c to be consistent with the experimental data analysis~\cite{J.Adam:2017}. The correlation structures of meson pairs are similar as those observed in $p$-Pb collisions at $\mathrm{\sqrt{s_{NN}}= 5.02}$ TeV as shown in  Figs.~\ref{pion_pair_correlation_pPb} and \ref{kaon_pair_correlation_pPb} but with the magnitudes of the near-side and away-side peaks significantly lower. 
This is first due to the many more uncorrelated particles in events at high multiplicity that decrease of the strength of the correlation functions, 
since for the case of no parton or hadron interactions (i.e. 
with $\sigma$ = 0 mb and $\mathrm{t_H}$ = 0 fm/c) these peaks in large systems are already much lower than those in small systems. 
It could also be because particles among the minijets are becoming more uniform in the azimuthal direction through interacting with a denser and hotter medium in Pb-Pb collisions. The clear difference is on the proton or anti-proton pair correlation functions. There is no near side depression in central and semi-central collisions in Pb-Pb collisions. Therefore these baryon-baryon correlations depend strongly on the multiplicity.

\begin{figure}[!htb]
	\centering
	\includegraphics[scale=0.35]{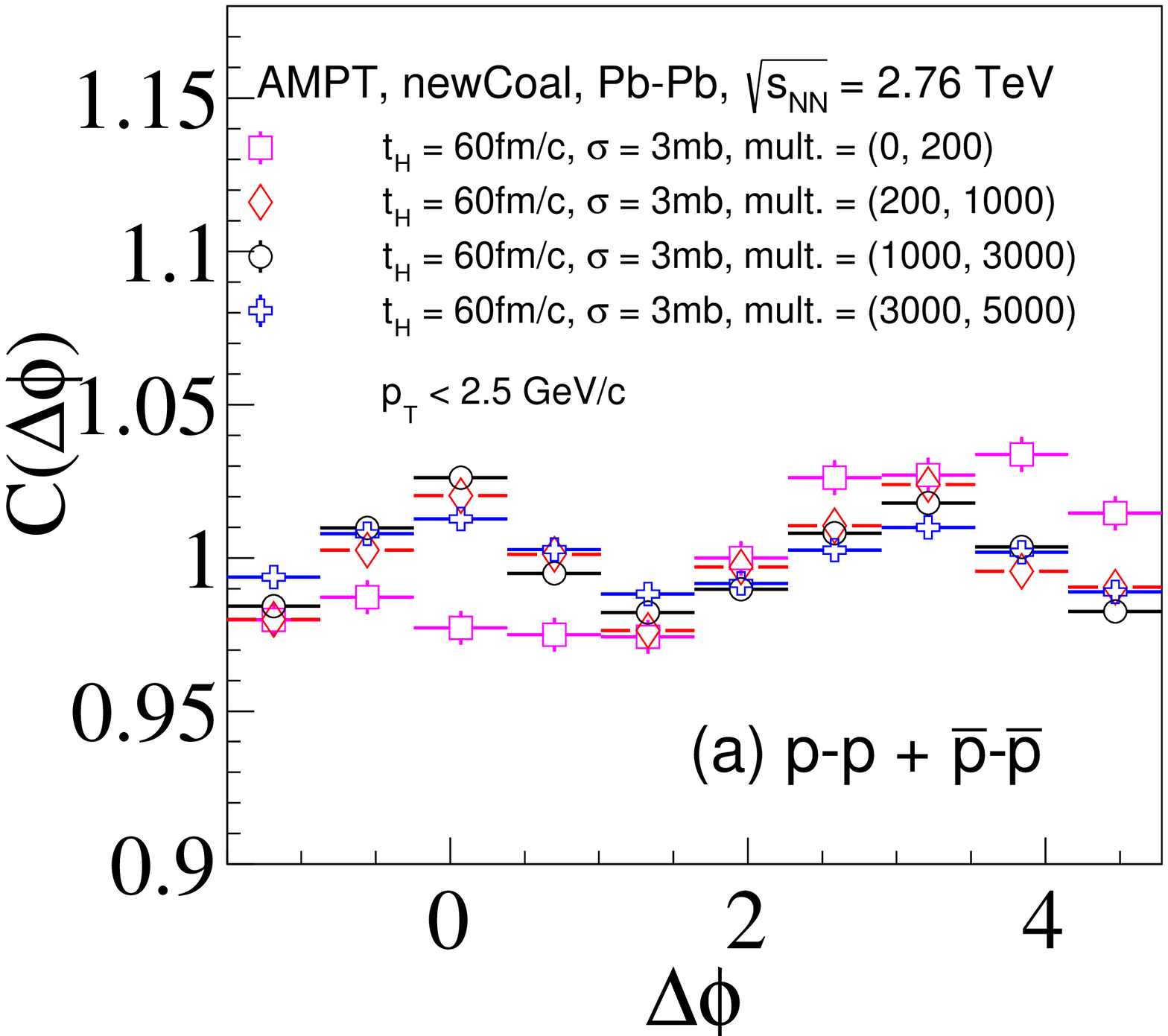}
	\includegraphics[scale=0.35]{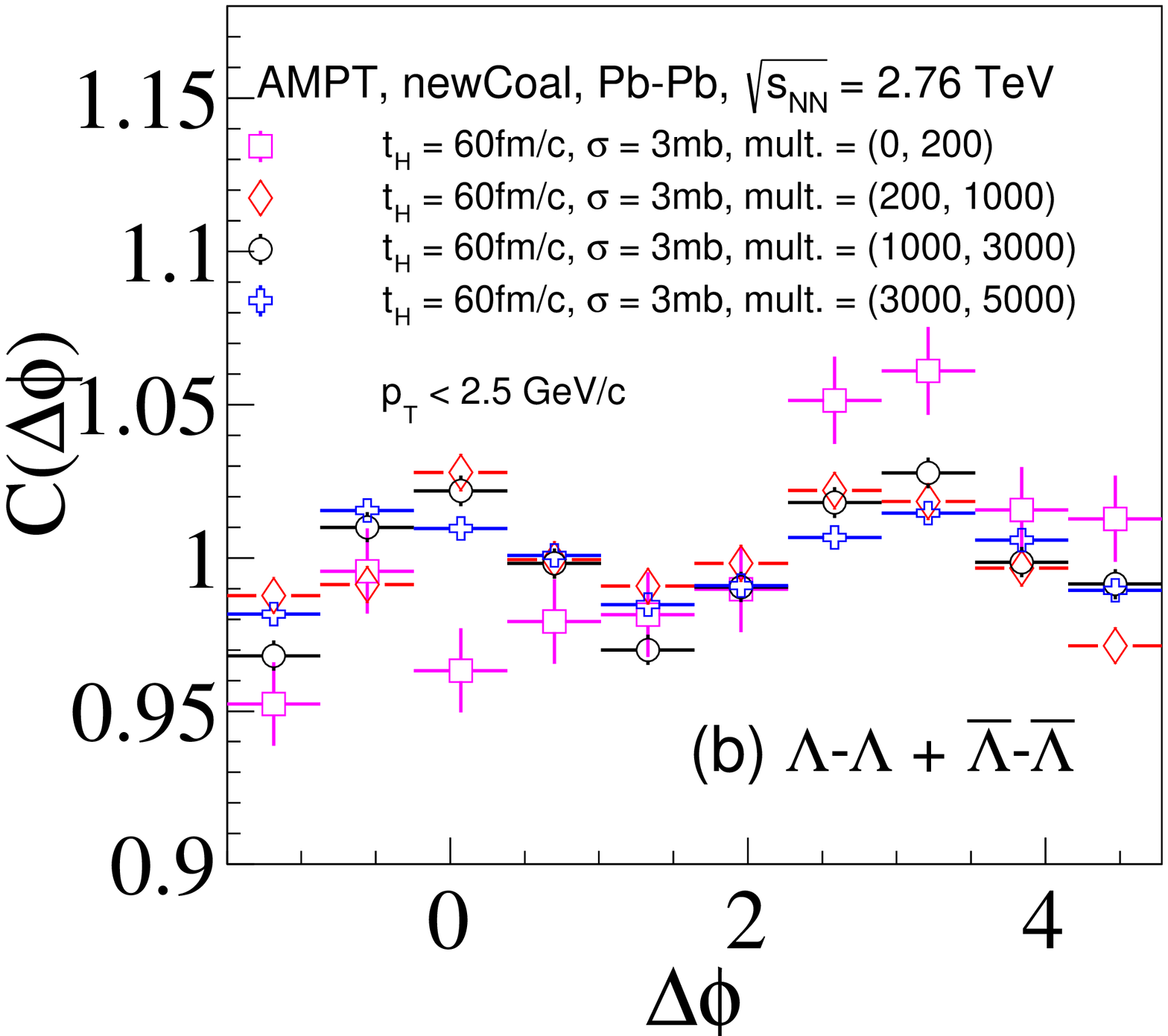}\\
	\includegraphics[scale=0.35]{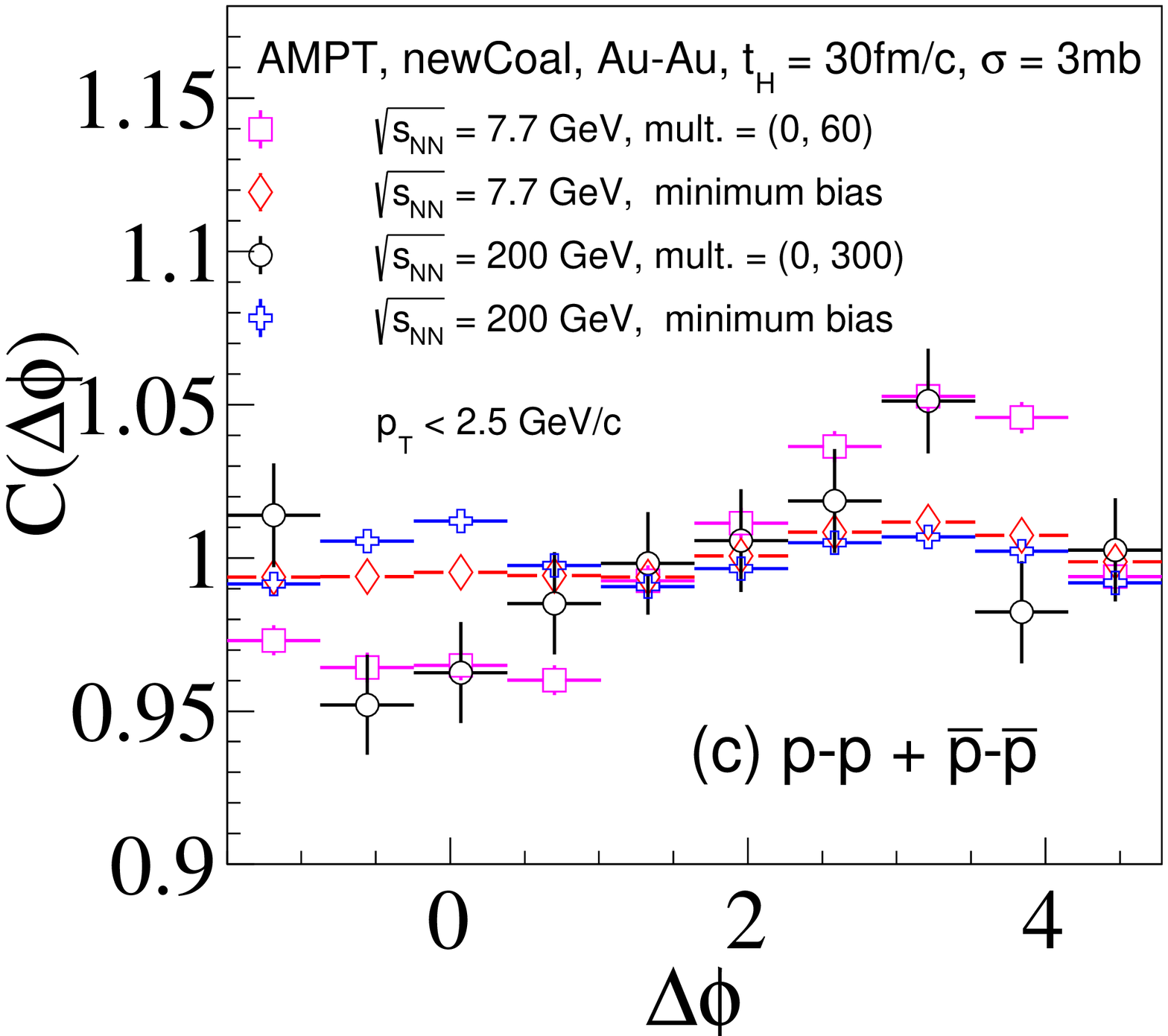}
	\includegraphics[scale=0.35]{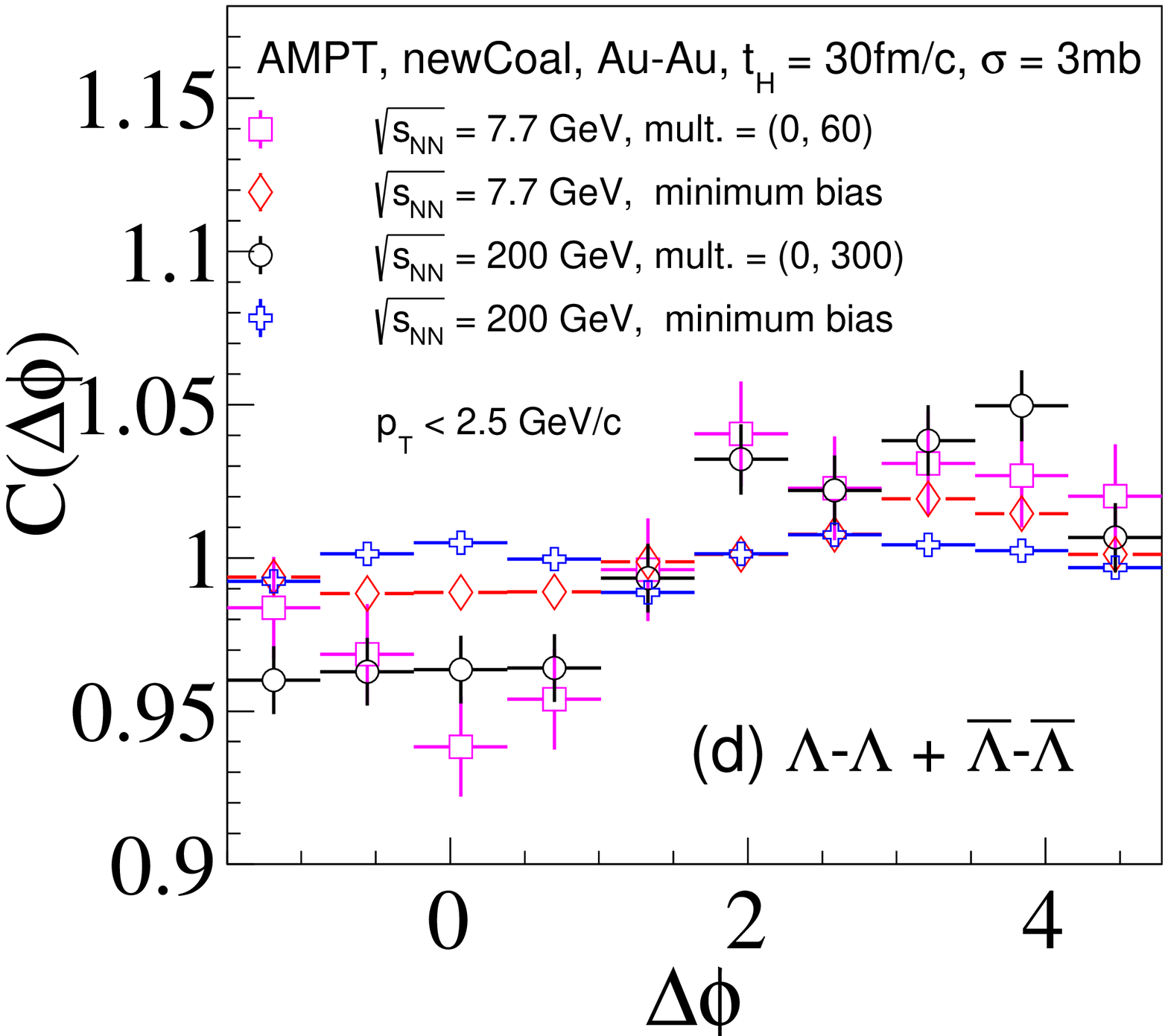}
	\caption{(Color online) One-dimensional $\Delta\phi$ correlation functions for proton-proton or $\Lambda$-$\Lambda$ pairs 
(including the corresponding antiparticle pairs) in Pb-Pb collisions at $\mathrm{\sqrt{s_{NN}}  = 2.76}$ TeV (upper panels) and in Au+Au collisions at $\mathrm{\sqrt{s_{NN}}}$ = 7.7 GeV and 200 GeV (lower panels) at different multiplicities from the string melting AMPT model with the new quark coalescence.
}
	\label{CFs_dependent_on_multiplicity}
\end{figure}

Results for more peripheral Pb-Pb collisions at $\mathrm{\sqrt{s_{NN}} = 2.76}$ TeV
on low-$p_T$ proton-proton ($\bar{p}$-$\bar{p}$) and $\Lambda$-$\Lambda$ ($\bar{\Lambda}$-$\bar{\Lambda}$) angular correlations are shown in 
Fig.~\ref{CFs_dependent_on_multiplicity}, 
together with the results for Au+Au collisions of different multiplicities at $\mathrm{\sqrt{s_{NN}}}$ = 7.7 GeV and 200  GeV. Usually we do not observe the near-side depression in Pb-Pb or Au+Au collisions, for example, in the  Pb-Pb results within multiplicity intervals of (200,1000), (1000,3000), (3000,5000) or in minimum bias Au+Au events. Here, multiplicity in Au-Au collisions is also defined as the number of charged tracks having $|y| <$ 0.8 and $p_{T}$ $<$ 2.5 GeV/c. Usually a clear near side peak is observed, similar to those for meson-meson or baryon-baryon pairs present in Figs.~\ref{pion_pair_correlation_PbPb}-\ref{proton_pair_correlation_PbPb}. 

\begin{figure}[!htb]
	\centering
	\includegraphics[scale=0.32]{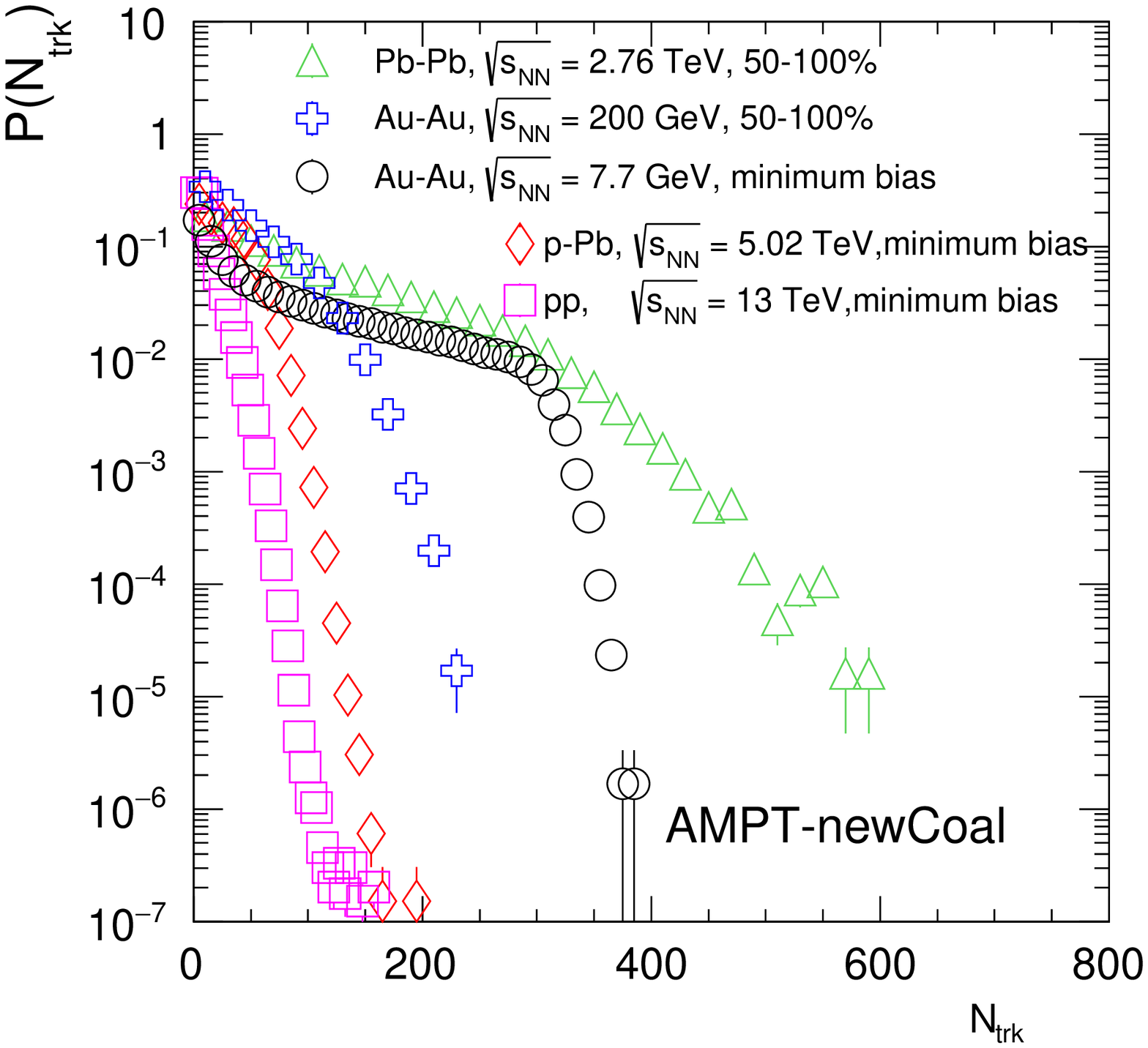}
	\caption{(Color online) Multiplicity ($\mathrm{N_{trk}}$) distributions for minimum bias $pp$ and $p$-Pb collisions at LHC energies and Au-Au collisions at RHIC energies as well as peripheral Pb-Pb collisions at LHC and Au-Au collisions at RHIC from the AMPT model with new quark coalescence.}  
	\label{CF_multi_dis}
\end{figure}

\begin{figure}[!htb]
	\centering
	\includegraphics[scale=0.35]{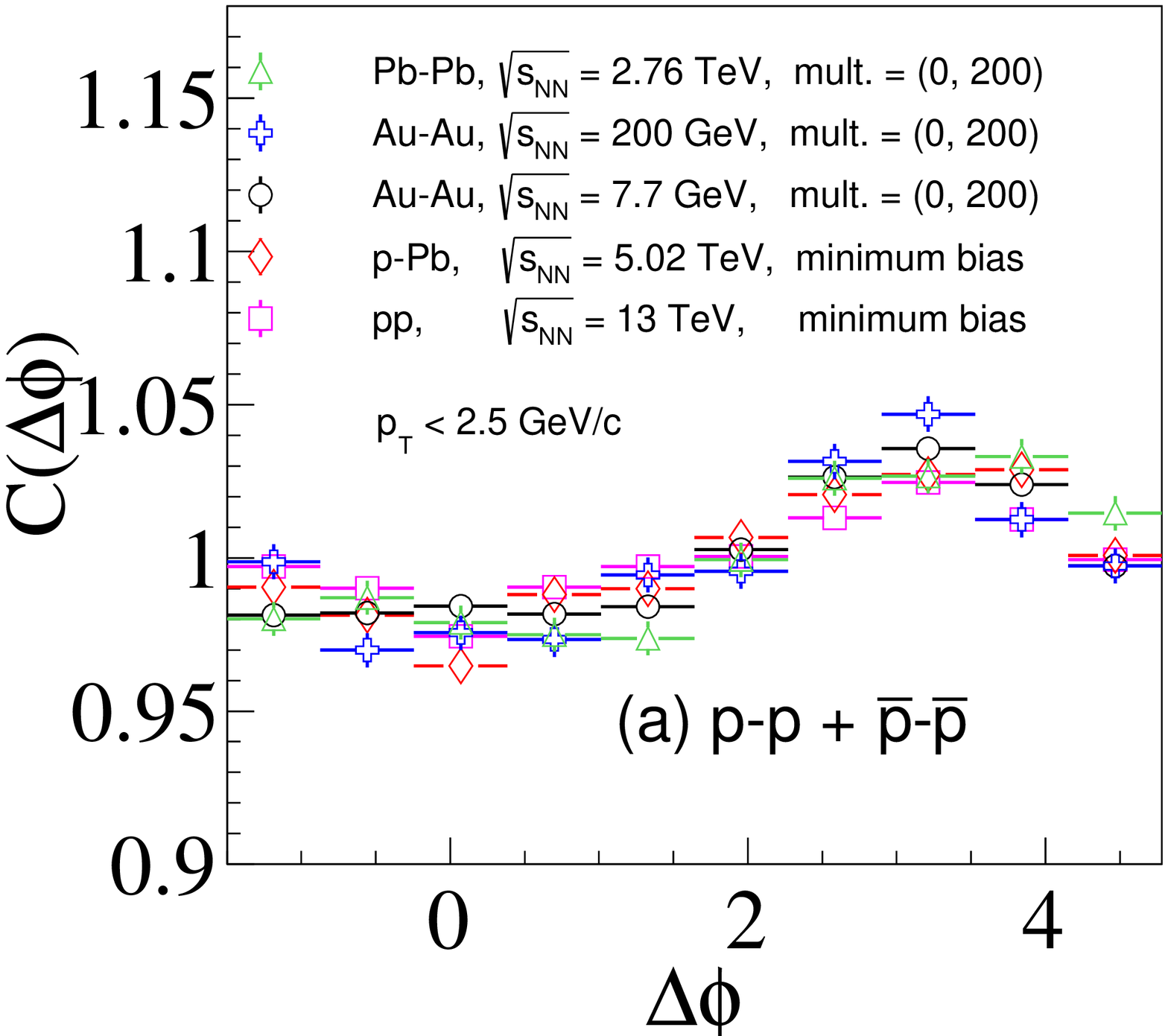}
	\includegraphics[scale=0.35]{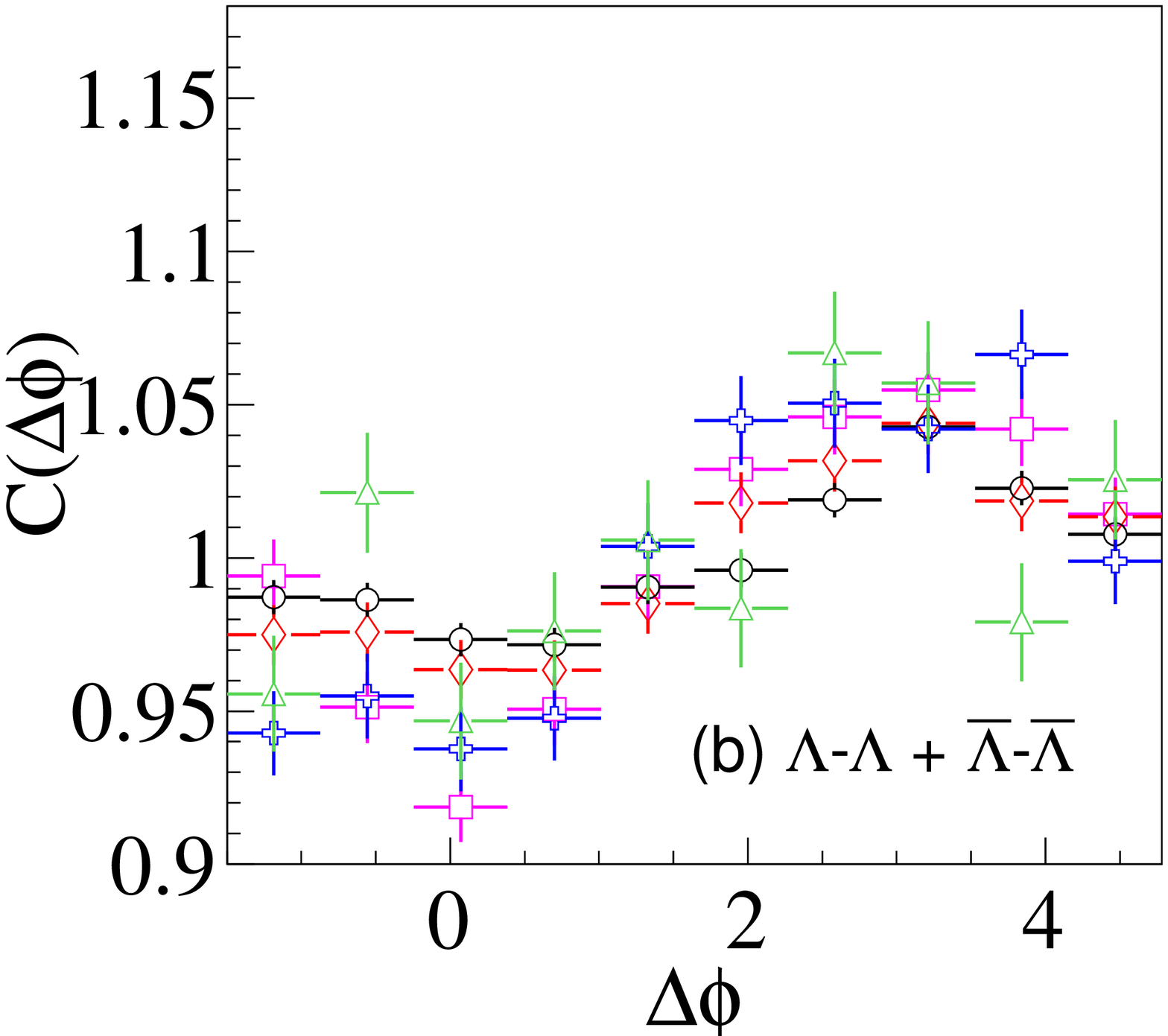}\\
	\caption{(Color online) One-dimensional $\Delta\phi$ correlation functions for proton-proton and  $\Lambda$-$\Lambda$ pairs in $pp$, $p$-Pb and Pb-Pb collisions at LHC energies and Au-Au collisions at RHIC energies within the similar multiplicity interval of (0-200) from the AMPT model with new quark coalescence.}
	\label{CFs_multi_to_diff_systems}
\end{figure}

However, we observe near-side depression in very peripheral Pb-Pb and Au-Au collisions in 
Fig.~\ref{CFs_dependent_on_multiplicity}, regardless of collisions energies. 
This multiplicity dependence of low $p_{T}$ $p$-$p$ ($\bar{p}$-$\bar{p}$) and $\Lambda$-$\Lambda$ ($\bar{\Lambda}$-$\bar{\Lambda}$) angular correlations can be further demonstrated through comparing the correlation functions across systems at similar multiplicity. Figure~\ref{CF_multi_dis} shows the multiplicity distributions for minimum bias collisions of $pp$ at $\mathrm{\sqrt{s}}$ = 13 TeV, $p$-Pb at $\mathrm{\sqrt{s_{NN}}}$ = 5.02 TeV and Au-Au at $\mathrm{\sqrt{s_{NN}}}$ = 7.7 GeV as well as for peripheral collisions (50-100$\%$) of Pb-Pb at $\mathrm{\sqrt{s_{NN}}}$ = 2.76 TeV and Au-Au at $\mathrm{\sqrt{s_{NN}}}$ = 200 GeV. Here, multiplicity is defined in the same way as done earlier. We see that the multiplicity distributions in different collision systems at LHC and RHIC energies share the range of (0,200). Thus we choose events with multiplicity within (0, 200) and compare their correlation functions in Fig.~\ref{CFs_multi_to_diff_systems}. It is rather surprising to see that the $p$-$p$ and $\Lambda$-$\Lambda$ angular correlations from these very different collision systems are very similar and all show the anti-correlation feature. This further demonstrates that the baryon pair anti-correlation is closely correlated to the event multiplicity.

\section{conclusion}
We have carried out a study on two-particle angular correlations for low $p_T$ pion, kaon, proton and $\Lambda$ pairs in $p$-Pb collisions at $\mathrm{\sqrt{s_{NN}}}$ = 5.02 TeV, Pb-Pb collisions at $\mathrm{\sqrt{s_{NN}}}$ = 2.76 TeV, and Au+Au collisions at $\mathrm{\sqrt{s_{NN}}}$ = 7.7 and 200 GeV with the string melting version of a multi-phase transport model with the new quark coalescence. Similar to our early study of $pp$ collisions at $\mathrm{\sqrt{s}}$ = 7 TeV~\cite{L.Y.Zhang:2018}, we find a clear depression in near-side baryon-baryon and antibaryon-antibaryon angular correlations in $p$-Pb collisions, and this depression feature is only present in Pb-Pb and Au-Au collisions at very low multiplicity. We also observe similar baryon pair correlations among $pp$, $p$-Pb, Pb-Pb collisions at LHC energies and Au-Au collisions at RHIC energies within a similar low multiplicity range. Comparisons of AMPT results with different parton cross sections and hadron cascade termination time indicate that parton interactions and the new quark coalescence are the key components to lead to the depression structure in baryon pair angular correlations. Our study shows that low-$p_T$ baryon-baryon anti-correlations at the near side, firstly observed in $pp$ collisions by the ALICE experiment, have a strong multiplicity dependence. Future analysis of the experimental data in heavy-ion collisions at LHC and RHIC will shed more light on the underlying physics of this phenomenon. 
 
 \section{acknowledgements}
Discussion with Prof. Jurgen Schukraft is greatly acknowledged. This work was supported in part by the Key Research Program of the Chinese Academy of Science with Grant No. XDPB09, by the Major State Basic Research Development Program in China under Contract No. 2015CB856904 and by the National Natural Science Foundation of China under Contract Nos. 11890710, 11775288, 11421505, 11628508 and 11520101004.

\end{document}